# Gate-Tunable Optical Anisotropy in Wafer-Scale, Aligned Carbon-Nanotube Films

Jason Lynch[1], Evan Smith[2], Adam Alfieri[1], Baokun Song[1, 3], Cindy Yueli Chen[4], Chavez Lawrence[1], Cherie Kagan[1, 4, 5], Honggang Gu[3], Shiyuan Liu[3], Lian-Mao Peng[6], Shivashankar Vangala[2], Joshua R. Hendrickson[2], and Deep Jariwala[1]

[1]Electrical and Systems Engineering, University of Pennsylvania, Philadelphia, PA 19104, US

[2]Air Force Research Laboratory, Sensors Directorate, Wright-Patterson Air Force Base, OH, 45433, USA

[3]State Key Laboratory of Digital Manufacturing Equipment and Technology, Huazhong University of Science and Technology, Wuhan 430074, China

[4]Department of Chemistry, University of Pennsylvania, Philadelphia, PA 19104, USA

[5]Department of Materials Science and Engineering, University of Pennsylvania, Philadelphia, PA 19104, USA

[6]Key Laboratory for the Physics and Chemistry of Nanodevices and Center for Carbon-based Electronics, School of Electronics, Peking University, Beijing, China

## Abstract

Telecommunications and polarimetry both require the active control of the polarization of light. Currently, this is done by combining intrinsically anisotropic materials with tunable isotropic materials into heterostructures using complicated fabrication techniques due to the lack of scalable materials that possess both properties. Tunable birefringent and dichromic materials are scarce and rarely available in high-quality thin-films over wafer scales. In this paper, we report semiconducting, highly aligned, single-walled carbon nanotubes (SWCNTs) over 4” wafers with normalized birefringence and dichroism values of 0.09 and 0.58, respectively. The real and imaginary part of the refractive index of these SWCNT films are tuned by up to 5.9% and 14.3% in the infrared at 2200 nm and 1660 nm, respectively, using electrostatic doping. Our results suggest that aligned SWCNTs are among the most anisotropic and tunable optical materials known and opens new avenues for their application in integrated photonics and telecommunications.

## Introduction

Active control of the polarization of light is critical for technologies such as optical communications, integrated photonics, and microscopy[1–3]. Although this is commonly done in the radio wave to terahertz regimes, actively controlling the polarization of light in the visible and near infrared remains difficult since few materials possess both optical anisotropy and highly-tunable optical properties. Liquid crystals are the most commonly used of these materials, but they require physical movement to tune their interaction with light. Not only does physical movement decrease their durability, but it also limits their switching speed which inhibits their incorporation into optical communication systems. Instead, materials whose optical properties can be tuned using external stimuli such as electric or magnetic fields are preferred in optical communication systems because of their fast-switching speed. However, the library of materials with tunable, in-plane optical anisotropy remains sparse. Even more scarce is the availability of such materials in high-quality thin films over wafer scales for practical applications.

Optical anisotropy is caused by structural asymmetries that result in the material having a complex refractive index ($\tilde{n} = n + ik$) that depends on the polarization of incident light. The asymmetry can occur as a result of either the intrinsic crystalline structure[4] or by etching/deposition anisotropic patterns in/on an isotropic material[5]. For uniaxial anisotropy, there are two directions that are symmetrical to one another, called the ordinary axes, that have the same refractive index ($\tilde{n}_{ord} = n_{ord} + ik_{ord}$). The third direction is the asymmetrical one and it is called the extraordinary axis, with a unique refractive index ($\tilde{n}_{ext} = n_{ext} + ik_{ext}$). The anisotropy is characterized by the normalized birefringence $\left(\frac{|n_{ord} - n_{ext}|}{n_{ord} + n_{ext}}\right)$ and normalized dichroism $\left(\frac{|k_{ord} - k_{ext}|}{k_{ord} + k_{ext}}\right)$. Materials such as transition metal dichalcogenides[6], perovskites[7], and polymers[8] exhibit optical anisotropy, but the extraordinary axis is typically out-of-plane which is less applicable to electro-optical systems since its effects can only be observed for light incident at extremely oblique angles far from normal incidence. Quantum confined materials such as black phosphorus[9], $ReS_2$[10], and one-dimensional crystals[11] all possess in-plane anisotropy. However, only a fraction of these materials have exhibited electrostatic tunability and optical anisotropy[12], and most cannot currently be grown over large areas, which limits their potential to be integrated into electro-optic systems at scale. Because of this, commercial electro-optic systems typically use bulky, anisotropic materials such as $LiNbO_3$, $BaTiO_3$, and $CaCO_3$. Quantum-confined materials, such as one-dimensional single-walled carbon nanotubes (SWCNTs), offer a unique opportunity to replace bulk crystals, enabling integrated, miniaturized, and energy efficient electro-optic systems.

While anisotropy enables polarization-dependent interactions with light, active polarization control requires that the material must also have optical properties that can be tuned using external stimuli such as strain[13], heat[14], electromagnetic fields[15–17], and electrostatic doping[18]. Electrostatic doping uses an applied voltage to shift the Fermi level which alters the carrier concentration and optical properties. Electrostatic gating is commonly used in telecommunications systems because of its efficient, fast switching behavior and relative ease of fabrication[19]. Although electrostatic doping effects can be observed in Si[20] and other bulk materials[21,22], the effect is most pronounced in low-dimensional materials. This is because the injected carriers accumulate near the surface of the material, and the optical properties are only tuned in this region.

SWCNTs are one-dimensional, seamless tubes of rolled single-layer graphene. The manner in which the SWCNTs are rolled is described by their chiral indices (n, m) which also determines their band structure. When mod(n-m) ≠ 0 and (2n+m) ≠ 0, the SWCNTs are semiconducting[23,24]; otherwise, they are metallic. The size of the band gap of semiconducting SWCNTs is inversely proportional to their diameter. Therefore, through the selection of the chiral indices, the band gap of SWCNTs can range from 0.1 to 1.8 eV[25]. The quantum confinement of carriers into one-dimensional tubes results in SWCNTs having excellent optical and electronic properties[26]. Unaligned systems of SWCNTs have shown tunable optical properties using electrostatic doping[27], and previous works have also shown optical anisotropy in bulk films of SWCNTs. However, these films were mixtures of metallic and semiconducting SWCNTs which results in unavoidable losses[28]. Recent advances in solution-phase separation and assembly approaches have enabled the production of wafer-scale, highly aligned, high-purity semiconducting (99.9999%) SWCNTs[29]. In this paper, we demonstrate giant gate-tunability in the optical anisotropy of these SWCNTs films. We also observed that aligned SWCNTs possess the largest normalized birefringence and dichroism in thin films (<4 nm) making them promising candidates to enable active control of light in the visible and near infrared regimes.

## Results and Discussion

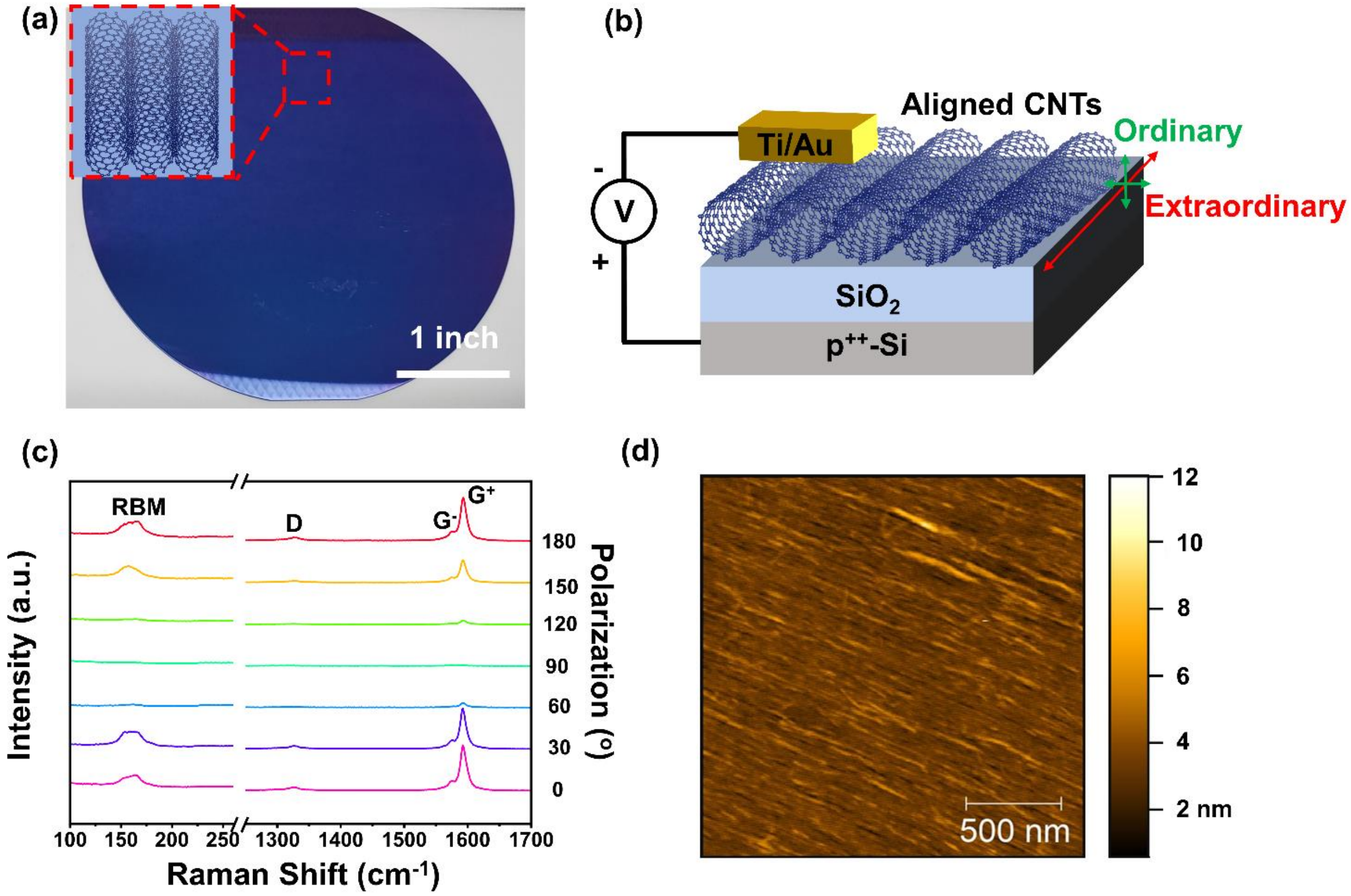

**Figure 1. Characterization of aligned SWCNT films (a)** Color photograph of a 4-inch wafer with globally aligned, high-purity (99.9999%) semiconducting SWCNTs prepared using a multiple dispersion and sorting process. The inset shows the direction of the global alignment. **(b)** Diagram of metal-oxide-semiconductor capacitor (MOSCap) geometry that enables the injection of holes and electrons into the SWCNT films by applying negative and positive voltages, respectively. The extraordinary axis (azimuthal direction) and ordinary axes (radial direction) of the SWCNTs are labelled. **(c)** Polarized Raman spectroscopy where the polarization angle is defined relative to the extraordinary axis of the SWCNTs. The radial breathing mode (RBM), D, and G modes are all labelled. The polarization dependence of the intensity of the $G^+$ mode gives an angular variation of 7.2°. **(d)** Atomic force microscopy (AFM) map of the surface of the SWCNT film showing the high degree of alignment and a surface roughness of 1.33 nm.

We used SWCNTs that were deposited using a multiple dispersion and sorting process that has been previously reported on to produce films that are high-density and 99.9999% semiconducting[29]. The fabrication process produced 4-inch wafers of globally-aligned SWCNT films, demonstrating that these films can be produced at scale (Figure 1a). However, the samples were cut into 1 cm x 1 cm squares for this study. The top electrical contact is 5 nm of Ti and 45 nm of Au that was deposited using electron-beam evaporation (See Methods) to make a metal-oxide-semiconductor capacitor (MOSCap) geometry (Figure 1b). In the MOSCap geometry, injected charge accumulates near the SWCNT dielectric interface, and the charge density drops off exponentially with a Debye length of ≈1 nm[30,31]. Despite the film thickness (3.7 nm) being larger than the Debye length, we approximated the refractive index as being constant throughout the film. The SWCNT film was grounded while the voltage was applied to the degenerately-doped Si substrate. Because of this, a positive (negative) voltage will decrease (increase) the Fermi level, resulting in an increase in the hole (electron) concentration.

The structural properties of the SWCNTs were studied using polarization-dependent Raman spectroscopy (Figure 1c). The radial breathing modes (RBM) with Raman shifts between 141 $cm^{-1}$ and 171 $cm^{-1}$ are the result of vibrations along the radial direction of the nanotubes. The Raman shift of the RBM is determined by the diameter of the SWCNTs[32] and the width of the RBM indicates the presence of multiple different diameters within the film. The two most abundant diameters are 1.46 nm and 1.57 nm (Supplemental Information Figure S1). The D-band mode (1323 $cm^{-1}$) is caused by the symmetry-breaking nature of defects within the SWCNTs showing that there are some defects present, but its low intensity relative to the G-band mode indicates a low concentration of defects[33]. The G-band mode comes from the first order resonance in bulk graphite samples, and it redshifts in SWCNTs due to their curvature and quantum confinement effects[34,35]. The G-band is split into two different modes called the $G^+$ and $G^-$ modes and the Raman shift difference of these two modes is consistent with past studies on semiconducting SWCNTs, further confirming the high quality of the films[34]. The RBM, D, and G modes all disappear in the Raman spectra as the polarization of the incident light is rotated which is consistent with all the SWCNTs being aligned in the same direction. By comparing the intensity of the $G^+$ mode for light polarized parallel and perpendicular to the extraordinary axis, the alignment variation of the SWCNTs was calculated to be $7.2^o$ [29]. Atomic force microscopy (AFM) further corroborates the highly-aligned nature of the films (Figure 1c). Using AFM, the surface roughness was calculated to be 1.33 nm which is significantly smaller than both the spot size (2 mm) and wavelength of light (600-2200 nm) that was used for spectroscopic ellipsometry.

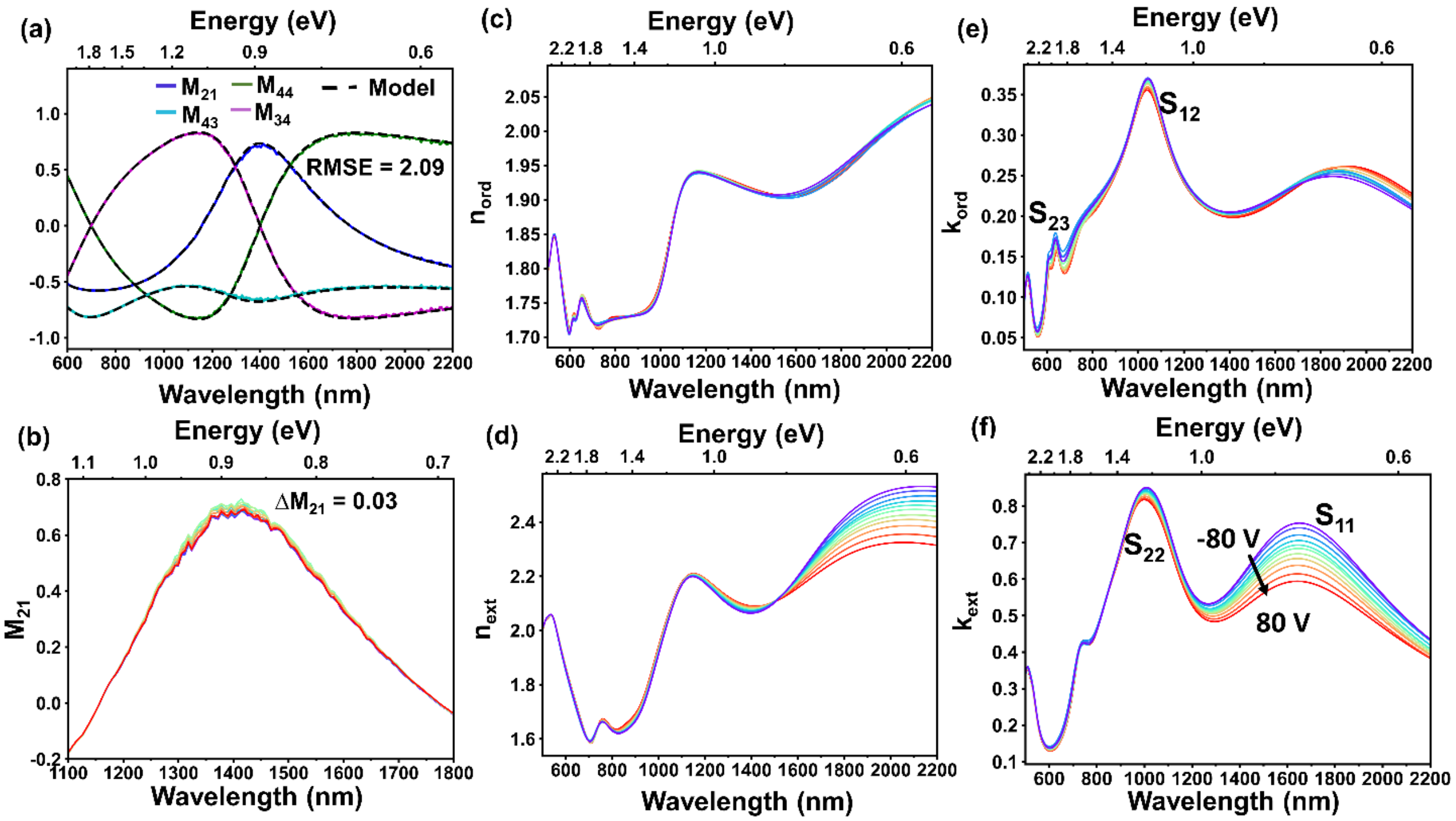

**Figure 2. Anisotropic, gate-tunable, complex refractive index of SWCNTs. (a)** Selected elements of the Mueller matrix measured at an incidence angle of 55° without any applied voltage. The model was fitted to the experimental data using a multi-Lorentz oscillator model and the root mean-squared error (RMSE) was 2.09. **(b)** The voltage dependence of the $M_{21}$ element which shows a larger tunability than the uncertainty of the measurement. The real part of the refractive index along the **(c)** ordinary and **(d)** extraordinary directions. The imaginary part of the refractive index along the **(e)** ordinary and **(f)** extraordinary directions with the zero change in angular momentum transitions ($S_{11}$ and $S_{22}$) and non-zero change in angular momentum transitions ($S_{12}$ and $S_{23}$) labelled.

The Mueller matrix is a 4x4 matrix that relates the Stokes vector of incident to reflected light, and it comprehensively describes the polarized light-matter interactions[36,37]. Selected elements of the Mueller matrix at an incident angle of 55° without an applied voltage are shown in Figure 2a. The normalized Mueller matrix was measured at 55°, 65°, and 75° angles relative to the normal direction for each voltage with the extraordinary axis of the SWCNTs at a 40° angle with the in-plane wavevector of the incident light (Supplementary Information S2). The $M_{21}$ element, which is directly related to the linear dichroic response, was found to change when a voltage was applied (Figure 2b), as well as other elements of the Mueller matrix (Supplementary Information Figure S2). The $M_{21}$ was found to change by up to 0.03 which is significantly larger than the instrumental uncertainty of ≈0.001 (Supplementary Information Figure S4). Therefore, the modulation in the Mueller matrix is due to a change in the refractive index of the SWCNTs instead of random variations between measurements. The anisotropic, complex refractive index was extracted from the Mueller matrix using a multi-Lorentz oscillator model[38] that minimized the root mean-squared error ($\mathrm{RMSE} = \sqrt{\frac{1}{15p-q}\sum_{i,j}\left(M_{ij}^{Mod} - M_{ij}^{Exp}\right)^2}$ where p is the number of wavelengths, q is the number of fit parameters, “Mod” and “Exp” denote the model and experimental values, and the sum is over all of Mueller matrix elements and wavelengths).

The extracted gate-tunable, complex refractive indices are shown in Figures 2c-f. The peaks around 1750 and 981 nm in the extraordinary direction are attributed to the $S_{11}$ and $S_{22}$ sub-band excitonic transitions, respectively. The $S_{11}$ peak location differs from the theoretical energies for 1.46 and 1.57 eV SWCNTs by 14 meV and 32 meV, respectively[39]. These peaks are broadened since they are the convolution of multiple chiralities of SWCNTs. The two fundamental transitions of $S_{12}$ and $S_{23}$ in the ordinary direction are 1.18 and 2.05 eV, respectively. A peak near the $S_{11}$ transition is also observed in the ordinary direction, but this is attributed to the fact the SWCNTs are not perfectly aligned allowing ordinarily polarized light to excite the $S_{11}$ transition in some of the nanotubes. The refractive index in the extraordinary direction is found to be larger than in the ordinary direction because optical selection rules result in the extraordinary direction resonances having larger oscillator strengths[40]. The refractive index can be tuned by injecting charges due to the plasma dispersion effect (PDE)[41]. The SWCNTs are naturally p-doped due to ambient adsorbates such as oxygen and water[42]. When a negative voltage is applied, electrons will be injected into the SWCNTs making the SWCNT become nearly intrinsic with minimal excess free carriers. The intrinsic nature of the SWCNTs reduces Pauli blocking[43] and increases the oscillator strength[44]. Conversely, when a positive voltage is applied, the SWCNTs become more heavily p-doped, resulting in increased Pauli blocking and decreased oscillator strength. Therefore, a positive (negative) voltage decreases (increases) the complex refractive index. The refractive index in the extraordinary direction has larger tunability than in the ordinary direction because its oscillator strength is more dependent on the exciton binding energy[40] which is consistent with previous measurements on the chemical doping effects of thin films of a mixture of metallic and semiconducting SWCNTs[45].

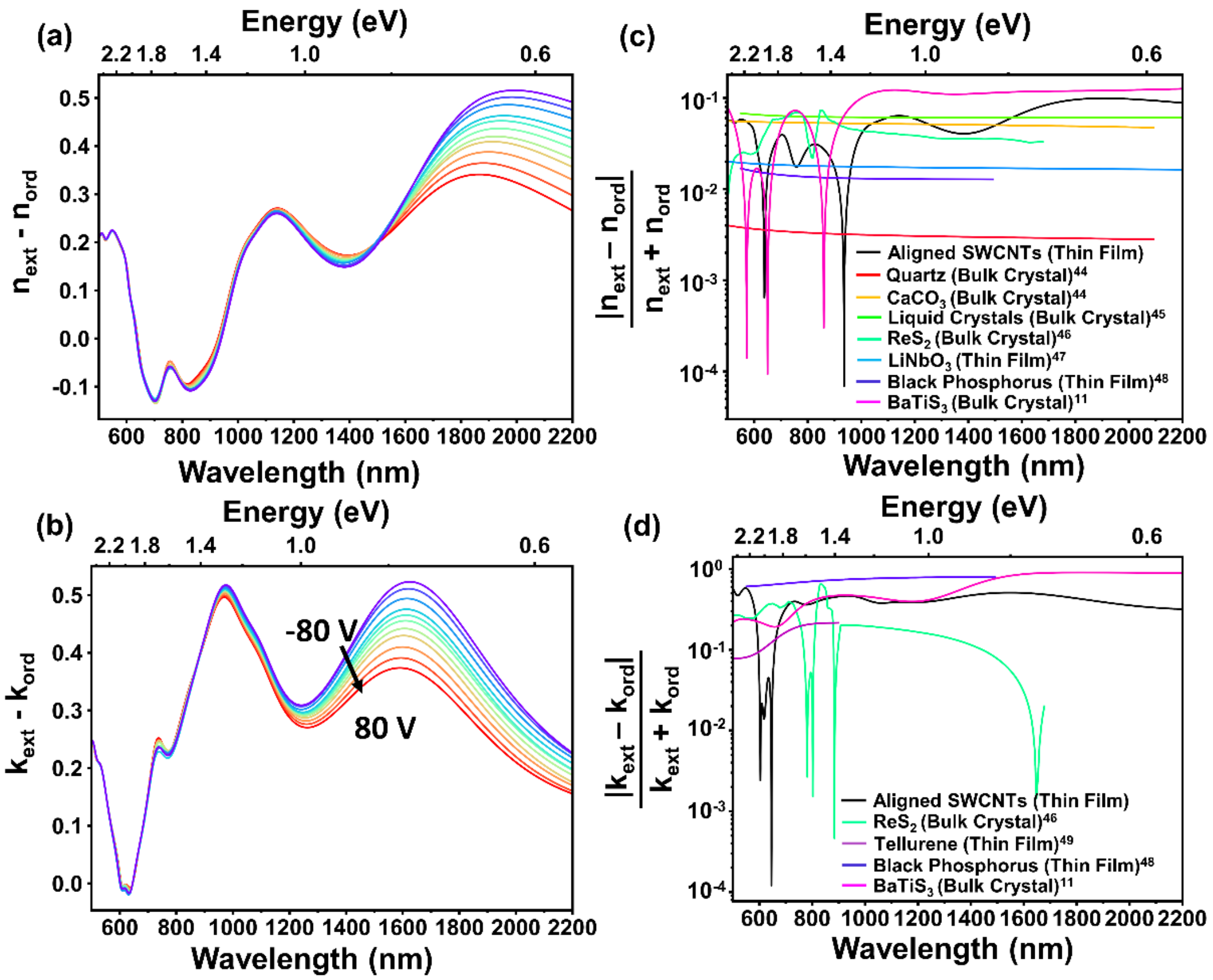

**Figure 3. Gate-tunable, in-plane birefringence and dichroism of SWCNTs.** The voltage dependence of the **(a)** birefringence ($n_{ext} - n_{ord}$) and **(b)** dichroism ($k_{ext} - k_{ord}$) in highly-aligned SWCNTs measured using Mueller matrix ellipsometry. The largest tunability in birefringence and dichroism is observed around the fundamental $S_{11}$ transition. The magnitude of the **(c)** normalized birefringence ($|n_{ext} - n_{ord}|/(n_{ext} + n_{ord})$) and **(d)** normalized dichroism ($|k_{ext} - k_{ord}|/(k_{ext} + k_{ord})$) in highly-aligned SWCNTs without an applied voltage are compared to other popular materials with in-plane anisotropy that were measured in either bulk crystal or thin film form[11,46–51].

The optical anisotropy in SWCNTs is caused by the optical selection rules where an electron can be excited by extraordinarily polarized light without needing a change in its angular momentum, whereas an electron that is excited by ordinarily polarized light must change its angular momentum. Aligned SWCNTs therefore possess both birefringence ($n_{ext} - n_{ord}$) and dichroism ($k_{ext} - k_{ord}$) because of these optical selection rules (Figures 3a and 3b). The depolarization effects of the optical selection rule in the ordinary directions results in dampening of the peaks which is why the birefringence and dichroism are both positive over the majority of the wavelengths studied. The optical selection rule leads to polarization-dependent optical bandgaps since the fundamental transition in the extraordinary and ordinary directions are the $S_{11}$ and $S_{12}$ transitions, respectively. Because of this, the normalized dichroism should approach unity for energies between these two transitions. However, the maximum value was only 58% since the SWCNTs are not perfectly aligned. Therefore, dichroism can be further improved by increasing the alignment of SWCNTs. Since the extraordinary refractive index has larger tunability than the ordinary refractive index, the tunability of the birefringence and dichroism is approximately equal to the tunability of the extraordinary refractive index. Therefore, the $S_{11}$ resonance showed the largest tunability for both birefringence and dichroism.

Aligned SWCNTs have normalized birefringence values in the near infrared that are comparable to well-known birefringent materials such as liquid crystals and $CaCO_3$ (Figure 3c). Aligned SWCNTs also show larger normalized birefringence values than other quantum-confined materials that are gate-tunable such as black phosphorus and $ReS_2$. SWCNTs show larger birefringence than these low-dimensional materials is that the optical selection rules for SWCNTs discussed above result in a relatively large optical bandgap difference between the extraordinary and ordinary axes. The optical bandgap difference for SWCNTs was calculated to be 436 meV while other low-dimensional materials do not possess similar rules. For example, $ReS_2$ has an optical bandgap difference of 72 meV between its two in-plane, perpendicular axes. Since the optical bandgap difference in SWCNTs is much larger than the linewidth of the $S_{12}$ resonance (210 meV), only the extraordinary direction is excitonic while the ordinary direction acts as a dielectric. To the best of our knowledge, this results in aligned SWCNTs having the largest normalized birefringence of any material < 4 nm thick making it an excellent candidate for compact, energy-efficient, active control of light. Aligned SWCNTs also display a large normalized dichroism value for the same reasons as its large birefringence (Figure 3d). However, its normalized dichroism is still lower than materials such as $BaTiS_3$ and black phosphorus whose dichroism are the result of crystalline bonds. As discussed above, the aligned SWCNTs normalized dichroism can be further increased by improved fabrication techniques since we still observed some $S_{11}$ absorption in the ordinary direction because the SWCNTs were not perfectly aligned. Further, it should also be noted that black phosphorus and $BaTiS_3$ are only available in bulk single crystal form where they exhibit this high normalized dichroism, limiting their potential applications. In addition, of all the materials compared, aligned SWCNTs possess the best combination of large, tunable anisotropy with ultra-thin, wafer-scale fabrication and ambient environment stability, making it ideal for numerous electro-optical applications.

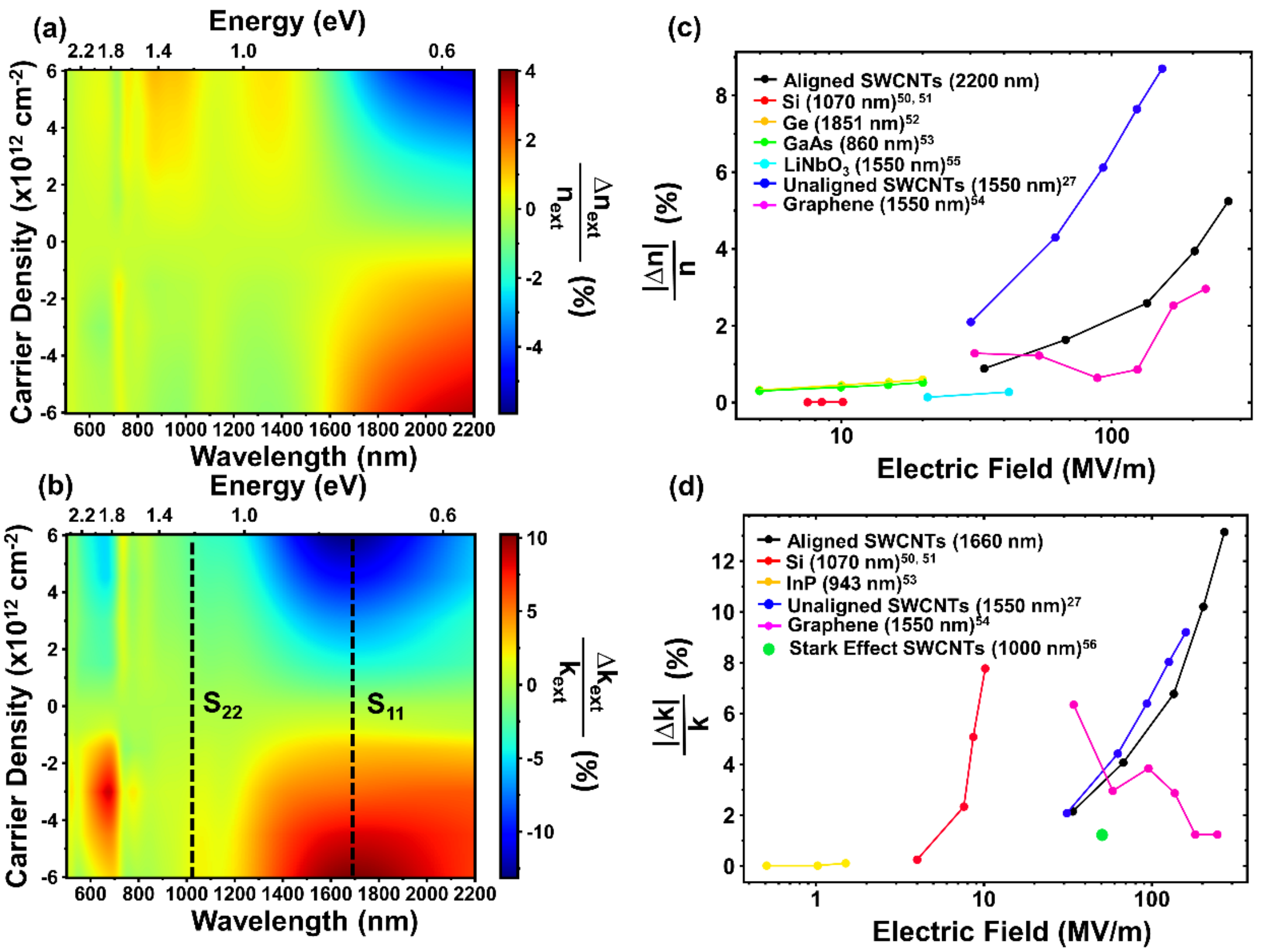

**Figure 4. Gate-tunability of the extraordinary axis of SWCNTs.** The percent change in **(a)** n and **(b)** k of the extraordinary axis of SWCNTs as a function of wavelength and carrier density with the two lowest order resonances ($S_{11}$ and $S_{22}$) labelled. Comparison of the tunability in the **(c)** normalized n and **(d)** normalized k of the extraordinary axis of SWCNTs with several other materials commonly used in the infrared regime where the dots represent the data points. The tunability values were taken from literature[27,52–58].

The normalized change in the real and imaginary parts of the refractive index for various wavelengths and carrier densities ($n_c$) are shown in Figures 4a and 4b, respectively. The carrier density was calculated using a parallel plate capacitor model, $n_c = \frac{\epsilon_0 \epsilon_{SiO2} V_g}{e\, d_{SiO2}}$ where $\epsilon_0$ is the vacuum permittivity, $\epsilon_{SiO2}$ is the relative permittivity of $SiO_2$ (3.9[59]), e is the charge of an electron, $V_g$ is the gate voltage, and $d_{SiO2}$ is the thickness of the $SiO_2$ (309 nm). The sign of $n_c$ also indicates the dominant carrier type where a positive (negative) carrier concentration means that the dominant carrier is holes (electrons). This model assumes that the intrinsic carrier density of the SWCNTs is much less than $n_c$ which is consistent with SWCNTs of this diameter[60]. Figure 4a and 4b also use a 2D interpolation to determine the tunability of the extraordinary refractive index between the carrier densities that were studied. The real and imaginary parts of the refractive index showed the largest tunability around the $S_{11}$ resonance since it has the largest oscillator strength of all the transitions when $V_g$ = 0V, making it most susceptible to PDE. The normalized change in the refractive index in the extraordinary direction is compared to other materials in Figures 4c and 4d. Bulk semiconductors (Si, Ge, and GaAs) all rely on bulk electro-optic effects, such as the Franz-Keldysh and Kerr effects, to tune their refractive index which enable broadband modulation of their complex refractive indices. However, the magnitude of the change in refractive index is relatively small. Because of the small change in refractive index, electro-optical systems based on these devices require interaction lengths that are typically much longer than the wavelength of light. Unlike these bulk semiconducting materials, the refractive index of SWCNTs, both aligned and unaligned, is dominated by its excitonic transitions. The primary exciton resonance ($S_{11}$) of SWCNTs enables it to achieve larger modulation than traditional bulk semiconductors and III-V multi-quantum wells (Figures 4c and 4d). In addition to the electrostatic driven PDE that we studied, the $S_{11}$ transition can also be modulated by the Stark effect[58]. The Stark effect uses an external electric field to shift the band gap of a material[61]. However, the Stark effect typically requires an electric field of similar strength to the ones we have studied here to shift the band gap on the order of several meV without significantly altering the intensity of the peak. Because of this small modulation, its effect is normally only seen at low temperatures, making it unfeasible for large-scale electro-optic applications. In addition, we also found electrostatic gating to yield a larger modulation in SWCNTs than the Stark effect as the imaginary part of the refractive index was modulated by 4.3% from electrostatic gating while it was only modulated by 1.2% by the Stark effect with an equivalent electric field intensity. However, it is worth noting that the Stark effect was measured in semiconducting (6, 5) chirality SWCNTs instead of the 1.46 nm diameter SWCNTs studied here. Therefore, electrostatically doped, highly-aligned SWCNTs can achieve larger modulation in its complex refractive index than traditional semiconductors since it is modulating a strong resonance, while electrostatic gating of SWCNTs outperforms the Stark effect in them.

## Conclusion

In conclusion, the tunability in the optical anisotropy of wafer-scale, highly-aligned, semiconducting SWCNTs was studied using Mueller matrix ellipsometry. Aligned SWCNTs were found to possess birefringence and dichroism comparable to the most widely used anisotropic materials such as $CaCO_3$ and liquid crystals in the near infrared and the normalized birefringence and dichroism in highly-aligned SWCNTs had record values for films <4 nm thick. However, unlike these traditional anisotropic materials, aligned SWCNTs were found to by tunable through electrostatic gating which will enable them to be implemented into active electro-optic systems without the need of it being incorporated into an heterostructure with an actively tunable material. The real and imaginary part of the refractive index of these SWCNT films are tuned by 5.9% and 14.3% in the infrared at 2200 nm and 1660 nm, respectively, which is larger than the tunability of Si, III-V semiconductors, and III-V multi-quantum wells. Our results have the potential to form the basis of ultra-thin, compact, and energy efficient electro-optical systems for the active control of the polarization of light.

## Methods

### SWCNT Film Fabrication

The highly-aligned SWCNT films were prepared using a previous multiple dispersion and sorting process[29]. The 0.3 cm x 0.3 cm metal contacts were deposited under vacuum using an E-beam evaporator (Kurt J. Lesker PVD 75). The 10 nm Ti and 90 nm Au were deposited sequentially, and wires were connected using an Ag paste. The sample was grounded using Ag paste to connect a wire to the $p^{++}$-silicon substrate.

### *Optical Characterization*

Raman spectroscopy was performed using a Horiba Scientific confocal microscope (LabRAM HR Evolution). This instrument is equipped with an Olympus objective lens (up to ×100) and three different grating (100, 600 and 1,800) -based spectrometers which are coupled to a Si focal plane array detector. A continuous-wave excitation source with excitation wavelength at 633 nm was used with the x100 objective lens at 3.2% laser power which corresponds to a power of 7 µW. The incident light passed through a linear polarizer before interacting with the sample.

### Mueller Matrix Ellipsometry

Spectroscopic ellipsometry was completed using a J.A. Woollam-RC2 ellipsometer in Mueller matrix mode which covers the visible and infrared wavelength regimes (300-2500 nm). Analysis of the data used the corresponding CompleteEase software to extract the anisotropic, complex optical constants (using a uniaxial, multi-Lorentz oscillator model) as well as film thickness and roughness. Ellipsometry was also performed on the bare $SiO_2$/Si substrate to improve the accuracy of our results.

## Acknowledgements

D. J. and J. L. acknowledge primary support for this work from the Office of Naval Research (N00014-23-1-203) and partial support from Northrop Grumman. J.R.H. acknowledges support from the Air Force Office of Scientific Research (Program Manager Dr. Gernot Pomrenke) under award number FA9550-20RYCOR059. L.P. acknowledges support from the National Science Foundation of China under grant number 61888102. H. G. and S. L. thank the Guangdong Basic and Applied Basic Research Foundation (No. 2023A1515030149). C.Y.C acknowledges support from the NSF Graduate Research Fellowship Program (NSF GRFP, DGE-1845298). C. K. and C. L. acknowledge support by the NSF (EEC-1941529).

# Supplementary Information: Gate-Tunable Optical Anisotropy in Wafer-Scale, Aligned Carbon-Nanotube Films

Jason Lynch[1], Evan Smith[2], Adam Alfieri[1], Baokun Song[1, 3], Cindy Yueli Chen[4], Chavez Lawrence[1], Cherie Kagan[1, 4, 5], Honggang Gu[3], Shiyuan Liu[3], Lian-Mao Peng[6], Shivashankar Vangala[2], Joshua R. Hendrickson[2], and Deep Jariwala[1]

[1]Electrical and Systems Engineering, University of Pennsylvania, Philadelphia, PA 19104, USA

[2]Air Force Research Laboratory, Sensors Directorate, Wright-Patterson Air Force Base, OH, 45433, USA

[3]State Key Laboratory of Digital Manufacturing Equipment and Technology, Huazhong University of Science and Technology, Wuhan 430074, China

[4]Department of Chemistry, University of Pennsylvania, Philadelphia, PA 19104, USA

[5]Department of Materials Science and Engineering, University of Pennsylvania, Philadelphia, PA 19104, USA

[6]Key Laboratory for the Physics and Chemistry of Nanodevices and Center for Carbon-based Electronics, Department of Electronics, Peking University, Beijing, China

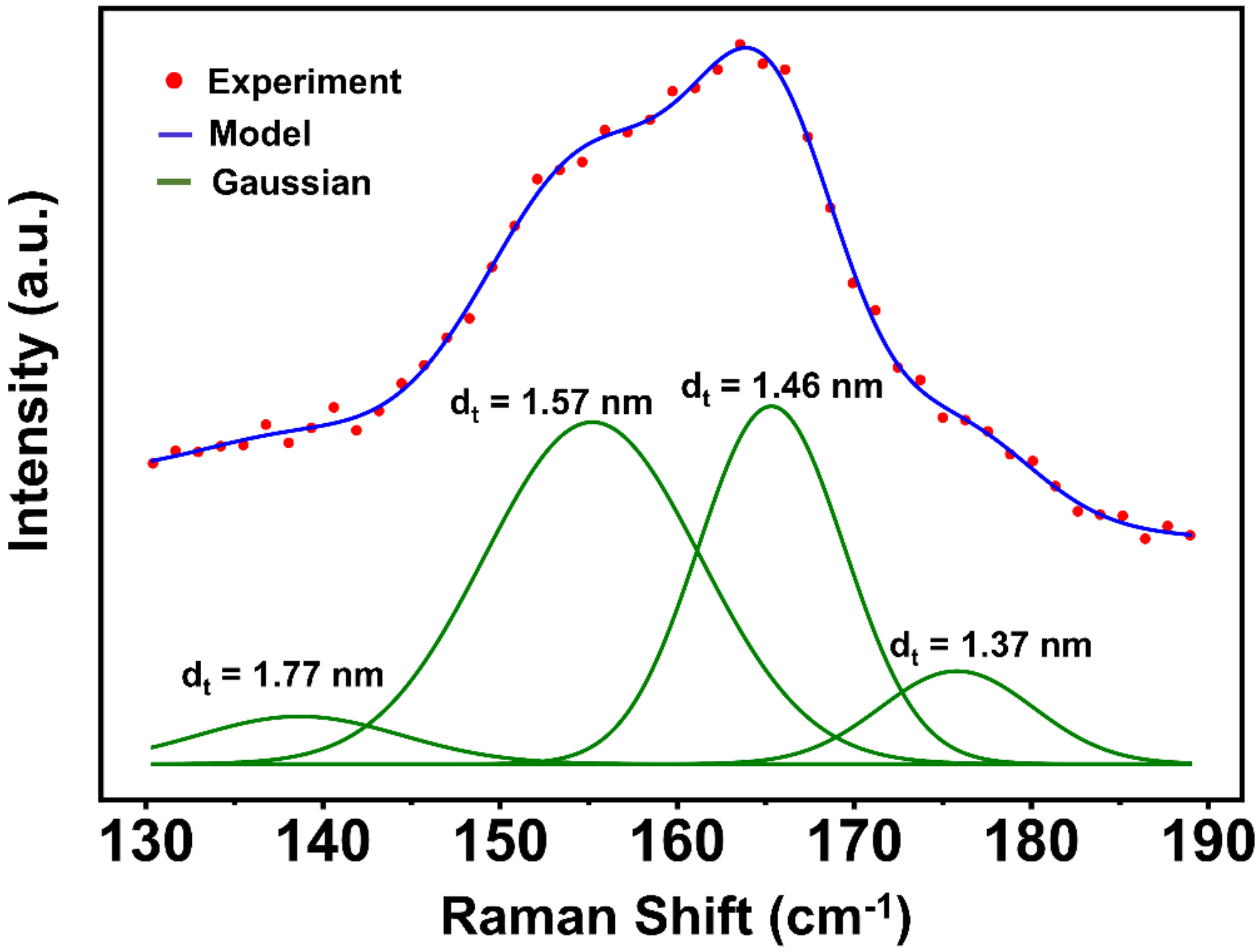

**Figure S1. Deconvolution radial breathing mode in the Raman Spectrum.** The Raman spectrum of aligned CNTs for light polarized parallel to the extraordinary axis was deconvoluted using a least square fit method to a series of Gaussian curves to identify the most common diameters and chiralities of carbon nanotubes.

## Mueller Matrix Ellipsometry

Standard ellipsometry measures the relative amplitude ($\psi = \arctan\left(\frac{|r_{TM}|}{|r_{TE}|}\right)$ where $r_{tm}$ and $r_{TE}$ are the reflection coefficients of the sample for TM and TE polarized light, respectively) and phase ($\Delta = \phi_{TM} - \phi_{TE}$ where $\phi_{TM}$ and $\phi_{TE}$ are the phases of TM and TE reflected light, respectively). Since there are two measurements at each wavelength for standard ellipsometry, it can accurately measure both the real and imaginary parts of the isotropic refractive index. However, for uniaxial, anisotropic materials, a minimum of four measurements at each wavelength is required to determine the complex refractive indices along the ordinary and extraordinary directions. Therefore, standard ellipsometry is insufficient for anisotropic. This lack of information for anisotropic materials can be overcome using multiple measurements of standard ellipsometry where the incident angle of light is varied when the extraordinary axis is out-of-plane, and the sample is rotated around its normal when the extraordinary axis is in-plane. However, a simpler approach for anisotropic materials is to use Mueller matrix ellipsometry.

The normalized Mueller matrix (M) relates the Stokes vector of the incident ($S_{inc}$) and reflected ($S_{ref}$) light ($S_{ref} = MS_{inc}$)[1,2]. The Stokes vector completely describes the polarization state of any wave of light, and it is defined as:

$$S = \begin{pmatrix} I_{Total} \\ I_{TM} - I_{TE} \\ I_{45} - I_{-45} \\ I_R - I_L \end{pmatrix} \quad (1)$$

where I is the intensity, and the subscripts describe the polarization (45 and -45 denote the cross polarizations and R and L denote right and left hand circularly polarizations, respectively). The Stokes vector is typically normalized such $I_{Total} = 1$. Since the Stokes vector spans polarization space, the Mueller matrix provides a comprehensive description of the light-matter interactions. Although $M_{11} = 1$ is fixed by the normalization process, Mueller matrix ellipsometry measures the 15 other Mueller matrix elements at every wavelength providing enough information to calculate the complex refractive index of anisotropic measurements.

The complex refractive index ($\tilde{n}$) of a semiconductor is modelled using a series of Lorentz oscillators[3]:

$$\tilde{n}^2(E) = \varepsilon_\infty + \sum_j \frac{f_j \Gamma_j E_j}{E_j^2 - E^2 + i\Gamma_j E_j} \quad (2)$$

where $\varepsilon_\infty$ is the static permittivity. The sum is over all of the resonant transitions of the semiconductor. $E_j$, $f_j$, and $\Gamma_j$ are the energy, oscillator strength, and linewidth of the $j^{th}$ resonance, and E is the energy of light. The models refractive index enables the Mueller matrix to be calculated using a transfer matrix method which is fitted to the experimental data by minimizing the root mean-squared error ( $RMSE =$ $\sqrt{\frac{1}{15p-q}\sum_i\left(\psi_i^{Mod}-\psi_i^{Exp}\right)^2+\left(\Delta_i^{Mod}-\Delta_i^{Exp}\right)^2}$ ) where p is the number of wavelengths, q is the number of fit parameters, "Mod" and "Exp" denote the model and experimental values, and the sum is over all of the wavelengths).

In order to ensure the accuracy of our model, we performed Mueller matrix ellipsometry at incident angles of 55°, 65°, and 75°. to reduce the effects of noise. We also aligned the extraordinary axis of the SWCNTs to a ≈35° with the in-plane wavevector of the incident light to observe the effects of the extraordinary and ordinary axes' refractive indices. Standard ellipsometry was also performed on the $SiO_2$/Si substrate without SWCNTs to determine the thickness of the $SiO_2$ layer. While fitting the Lorentz oscillator model to the experimental data, we used the CompleteEase software from J.A. Woollam, and the fit parameters were the Lorentz oscillators, the static permittivities, SWCNT layer thickness, and in-plane alignment of the extraordinary axis. The SWCNTs layer thickness and in-plane alignment of the extraordinary axis were determined using the zero voltage measurements, and they were held constant for all other voltages. Our models resulted in RMSE values < 3 which indicates their accuracy.

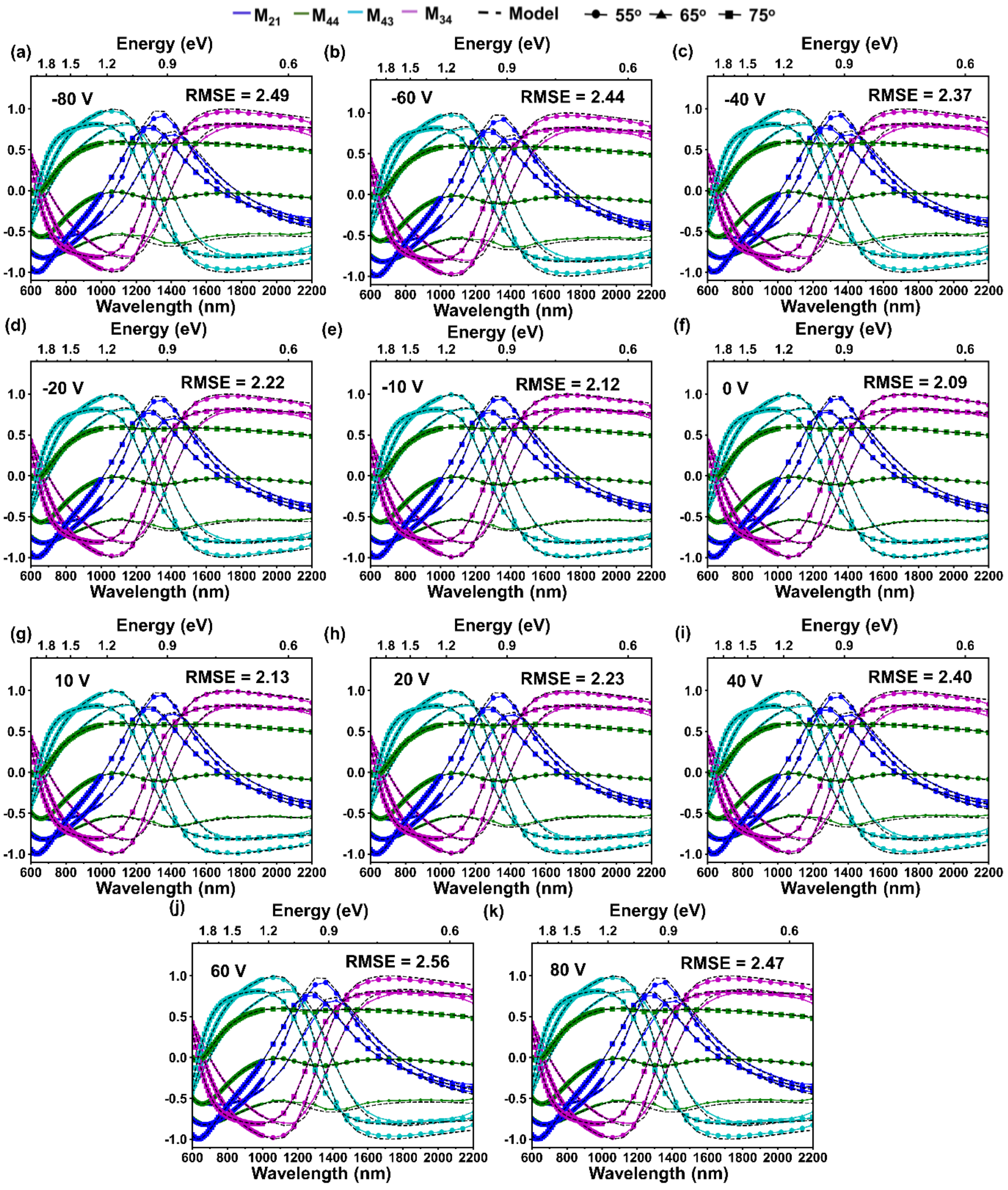

**Figure S2. Voltage dependence of select Mueller matrix elements.** Experimental and model values of select Mueller matrix elements when the extraordinary axis of the CNTs makes a 40° degree angle with the wavevector of the incident light at voltages of **(a)** -80 V, **(b)** -60 V, **(c)** -40 V, **(d)** -20 V, **(e)** -10 V, **(f)** 0 V, **(g)** 10 V, **(h)** 20 V, **(i)** 40 V, **(j)** 60 V, and **(k)** 80 V with the root-mean-square-error (RMSE) values for each voltage listed.

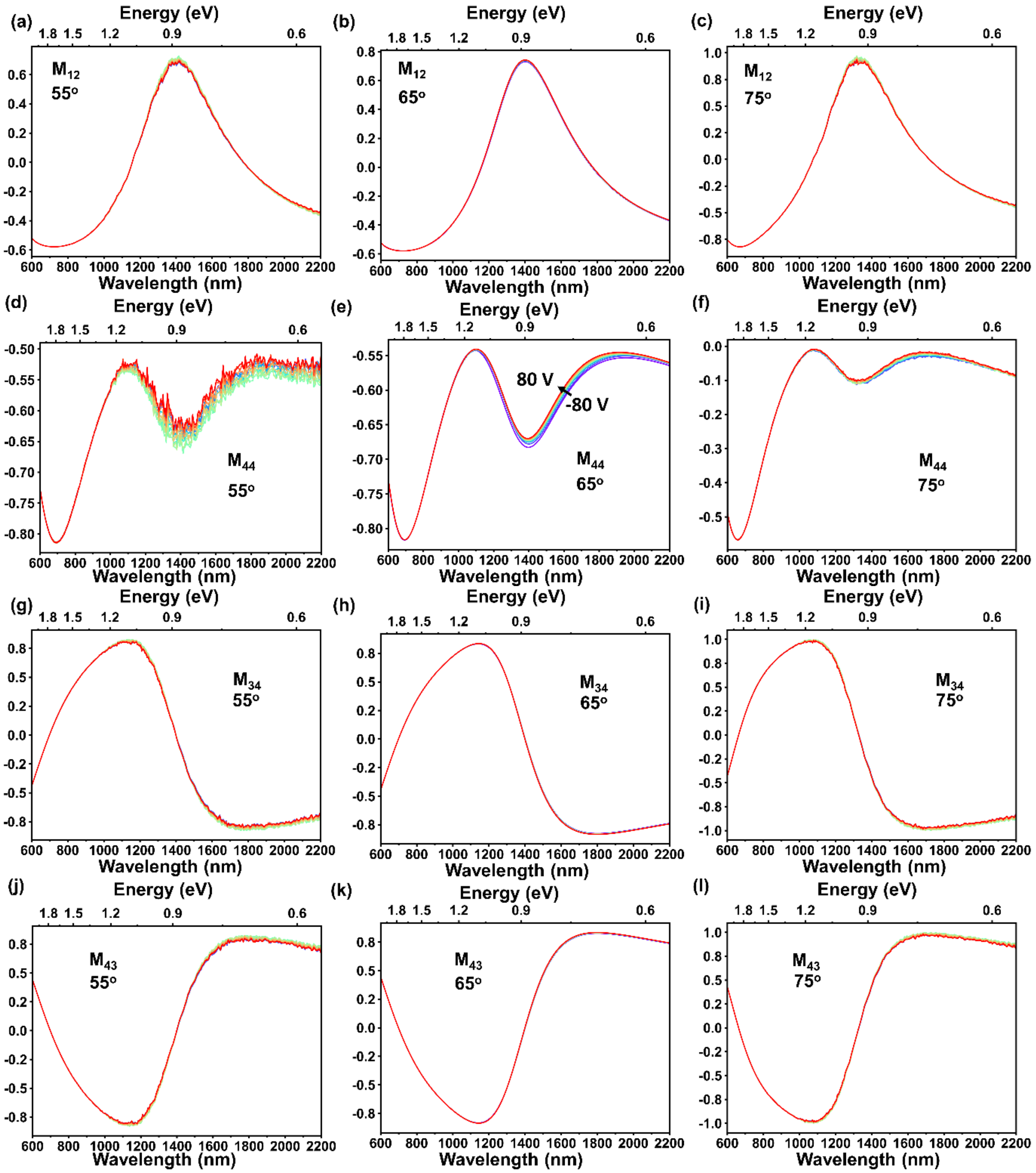

**Figure S3. Gate dependence of select Mueller matrix elements.** The gate dependence of $M_{12}$ at angles of incidence of **(a)** 55°, **(b)** 65°, and **(c)** 75°, $M_{44}$ at angles of incidence of **(d)** 55°, **(e)** 65°, and **(f)** 75°, $M_{34}$ at angles of incidence of **(g)** 55°, **(h)** 65°, and **(i)** 75°, $M_{43}$ at angles of incidence of **(j)** 55°, **(k)** 65°, and **(l)** 75°.

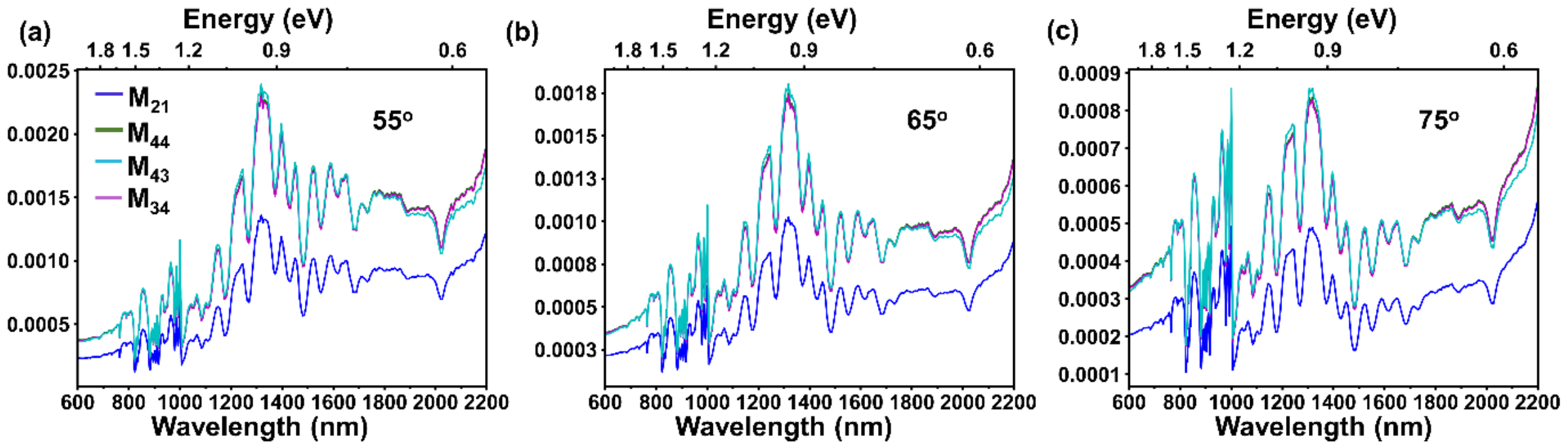

**Figure S4. Error of Mueller matrix elements.** The uncertainty of select Mueller matrix elements without an applied voltage at incident angles of **(a)** 55°, **(b)** 65°, and **(c)** 75°.

**Table S1.** Anisotropic, complex refractive index of aligned SWCNTs for gate voltages of -80 V, 0 V, and 80 V on an $SiO_2$/Si substrate

| | V = -80 V | | | | V = 0 V | | | | V = 80 V | | | |
|---|---|---|---|---|---|---|---|---|---|---|---|---|
| Wavelength (nm) | $n_{ord}$ | $k_{ord}$ | $n_{ext}$ | $k_{ext}$ | $n_{ord}$ | $k_{ord}$ | $n_{ext}$ | $k_{ext}$ | $n_{ord}$ | $k_{ord}$ | $n_{ext}$ | $k_{ext}$ |
| 500 | 1.79 | 0.10 | 2.00 | 0.34 | 1.80 | 0.09 | 2.01 | 0.34 | 1.80 | 0.09 | 2.01 | 0.34 |
| 501 | 1.79 | 0.10 | 2.00 | 0.35 | 1.80 | 0.09 | 2.01 | 0.34 | 1.80 | 0.10 | 2.01 | 0.34 |
| 502 | 1.80 | 0.10 | 2.01 | 0.35 | 1.80 | 0.10 | 2.01 | 0.34 | 1.80 | 0.10 | 2.01 | 0.35 |
| 503 | 1.80 | 0.11 | 2.01 | 0.35 | 1.80 | 0.10 | 2.01 | 0.35 | 1.80 | 0.10 | 2.01 | 0.35 |
| 504 | 1.80 | 0.11 | 2.01 | 0.35 | 1.80 | 0.10 | 2.02 | 0.35 | 1.80 | 0.11 | 2.02 | 0.35 |
| 505 | 1.80 | 0.11 | 2.01 | 0.36 | 1.80 | 0.11 | 2.02 | 0.35 | 1.80 | 0.11 | 2.02 | 0.36 |
| 506 | 1.80 | 0.11 | 2.02 | 0.36 | 1.80 | 0.11 | 2.02 | 0.36 | 1.81 | 0.11 | 2.02 | 0.36 |
| 507 | 1.80 | 0.12 | 2.02 | 0.36 | 1.81 | 0.11 | 2.02 | 0.36 | 1.81 | 0.12 | 2.02 | 0.36 |
| 508 | 1.81 | 0.12 | 2.02 | 0.36 | 1.81 | 0.12 | 2.03 | 0.36 | 1.81 | 0.12 | 2.03 | 0.36 |
| 509 | 1.81 | 0.12 | 2.03 | 0.35 | 1.81 | 0.12 | 2.03 | 0.36 | 1.81 | 0.12 | 2.03 | 0.36 |
| 510 | 1.81 | 0.12 | 2.03 | 0.35 | 1.81 | 0.12 | 2.03 | 0.36 | 1.81 | 0.12 | 2.03 | 0.36 |
| 511 | 1.81 | 0.12 | 2.03 | 0.35 | 1.82 | 0.12 | 2.03 | 0.36 | 1.82 | 0.12 | 2.03 | 0.36 |
| 512 | 1.82 | 0.12 | 2.03 | 0.35 | 1.82 | 0.12 | 2.04 | 0.35 | 1.82 | 0.12 | 2.04 | 0.36 |
| 513 | 1.82 | 0.12 | 2.04 | 0.35 | 1.82 | 0.12 | 2.04 | 0.35 | 1.82 | 0.13 | 2.04 | 0.35 |
| 514 | 1.82 | 0.12 | 2.04 | 0.34 | 1.82 | 0.12 | 2.04 | 0.35 | 1.82 | 0.13 | 2.04 | 0.35 |
| 515 | 1.82 | 0.12 | 2.04 | 0.34 | 1.83 | 0.12 | 2.04 | 0.35 | 1.83 | 0.13 | 2.04 | 0.35 |
| 516 | 1.83 | 0.12 | 2.04 | 0.34 | 1.83 | 0.12 | 2.04 | 0.35 | 1.83 | 0.13 | 2.04 | 0.35 |
| 517 | 1.83 | 0.12 | 2.04 | 0.34 | 1.83 | 0.12 | 2.05 | 0.34 | 1.83 | 0.13 | 2.05 | 0.35 |
| 518 | 1.83 | 0.12 | 2.04 | 0.33 | 1.83 | 0.12 | 2.05 | 0.34 | 1.83 | 0.13 | 2.05 | 0.34 |
| 519 | 1.83 | 0.12 | 2.04 | 0.33 | 1.84 | 0.12 | 2.05 | 0.34 | 1.84 | 0.12 | 2.05 | 0.34 |
| 520 | 1.84 | 0.12 | 2.04 | 0.33 | 1.84 | 0.12 | 2.05 | 0.34 | 1.84 | 0.12 | 2.05 | 0.34 |
| 521 | 1.84 | 0.11 | 2.04 | 0.33 | 1.84 | 0.12 | 2.05 | 0.33 | 1.84 | 0.12 | 2.05 | 0.34 |
| 522 | 1.84 | 0.11 | 2.05 | 0.32 | 1.84 | 0.12 | 2.05 | 0.33 | 1.84 | 0.12 | 2.05 | 0.33 |
| 523 | 1.84 | 0.11 | 2.05 | 0.32 | 1.84 | 0.12 | 2.05 | 0.33 | 1.84 | 0.12 | 2.05 | 0.33 |
| 524 | 1.84 | 0.11 | 2.05 | 0.32 | 1.84 | 0.11 | 2.05 | 0.32 | 1.84 | 0.12 | 2.05 | 0.33 |
| 525 | 1.84 | 0.11 | 2.05 | 0.32 | 1.85 | 0.11 | 2.05 | 0.32 | 1.85 | 0.11 | 2.05 | 0.32 |
| 526 | 1.84 | 0.10 | 2.05 | 0.31 | 1.85 | 0.11 | 2.05 | 0.32 | 1.85 | 0.11 | 2.05 | 0.32 |
| 527 | 1.85 | 0.10 | 2.05 | 0.31 | 1.85 | 0.11 | 2.05 | 0.32 | 1.85 | 0.11 | 2.05 | 0.32 |
| 528 | 1.85 | 0.10 | 2.05 | 0.31 | 1.85 | 0.10 | 2.05 | 0.31 | 1.85 | 0.11 | 2.06 | 0.31 |
| 529 | 1.85 | 0.10 | 2.05 | 0.31 | 1.85 | 0.10 | 2.05 | 0.31 | 1.85 | 0.10 | 2.06 | 0.31 |
| 530 | 1.85 | 0.09 | 2.05 | 0.30 | 1.85 | 0.10 | 2.06 | 0.31 | 1.85 | 0.10 | 2.06 | 0.31 |
| 531 | 1.85 | 0.09 | 2.05 | 0.30 | 1.85 | 0.10 | 2.06 | 0.30 | 1.85 | 0.10 | 2.06 | 0.30 |
| 532 | 1.85 | 0.09 | 2.05 | 0.30 | 1.85 | 0.09 | 2.06 | 0.30 | 1.85 | 0.09 | 2.06 | 0.30 |
| 533 | 1.84 | 0.09 | 2.06 | 0.29 | 1.85 | 0.09 | 2.06 | 0.30 | 1.85 | 0.09 | 2.06 | 0.30 |
| 534 | 1.84 | 0.08 | 2.06 | 0.29 | 1.85 | 0.09 | 2.06 | 0.29 | 1.85 | 0.09 | 2.06 | 0.29 |
| 535 | 1.84 | 0.08 | 2.06 | 0.28 | 1.84 | 0.09 | 2.06 | 0.29 | 1.84 | 0.09 | 2.06 | 0.29 |
| 536 | 1.84 | 0.08 | 2.06 | 0.28 | 1.84 | 0.08 | 2.06 | 0.28 | 1.84 | 0.08 | 2.06 | 0.28 |
| 537 | 1.84 | 0.08 | 2.06 | 0.27 | 1.84 | 0.08 | 2.06 | 0.28 | 1.84 | 0.08 | 2.06 | 0.28 |
| 538 | 1.84 | 0.07 | 2.06 | 0.27 | 1.84 | 0.08 | 2.06 | 0.27 | 1.84 | 0.08 | 2.06 | 0.27 |
| 539 | 1.84 | 0.07 | 2.06 | 0.27 | 1.84 | 0.08 | 2.06 | 0.27 | 1.84 | 0.08 | 2.06 | 0.27 |
| 540 | 1.84 | 0.07 | 2.06 | 0.26 | 1.84 | 0.07 | 2.06 | 0.26 | 1.84 | 0.07 | 2.06 | 0.26 |
| 541 | 1.83 | 0.07 | 2.06 | 0.26 | 1.84 | 0.07 | 2.06 | 0.26 | 1.83 | 0.07 | 2.06 | 0.26 |
| 542 | 1.83 | 0.07 | 2.05 | 0.25 | 1.83 | 0.07 | 2.05 | 0.26 | 1.83 | 0.07 | 2.05 | 0.26 |
| 543 | 1.83 | 0.06 | 2.05 | 0.25 | 1.83 | 0.07 | 2.05 | 0.25 | 1.83 | 0.07 | 2.05 | 0.25 |
| 544 | 1.83 | 0.06 | 2.05 | 0.24 | 1.83 | 0.07 | 2.05 | 0.25 | 1.83 | 0.07 | 2.05 | 0.25 |
| 545 | 1.83 | 0.06 | 2.05 | 0.24 | 1.83 | 0.06 | 2.05 | 0.24 | 1.82 | 0.07 | 2.05 | 0.24 |
| 546 | 1.82 | 0.06 | 2.05 | 0.23 | 1.82 | 0.06 | 2.05 | 0.24 | 1.82 | 0.06 | 2.05 | 0.24 |
| 547 | 1.82 | 0.06 | 2.05 | 0.23 | 1.82 | 0.06 | 2.04 | 0.23 | 1.82 | 0.06 | 2.04 | 0.23 |
| 548 | 1.82 | 0.06 | 2.04 | 0.22 | 1.82 | 0.06 | 2.04 | 0.23 | 1.82 | 0.06 | 2.04 | 0.23 |
| 549 | 1.82 | 0.06 | 2.04 | 0.22 | 1.82 | 0.06 | 2.04 | 0.22 | 1.81 | 0.06 | 2.04 | 0.22 |
| 550 | 1.81 | 0.06 | 2.04 | 0.21 | 1.81 | 0.06 | 2.04 | 0.22 | 1.81 | 0.06 | 2.04 | 0.22 |
| 551 | 1.81 | 0.05 | 2.04 | 0.21 | 1.81 | 0.06 | 2.03 | 0.22 | 1.81 | 0.06 | 2.03 | 0.21 |
| 552 | 1.81 | 0.05 | 2.03 | 0.21 | 1.81 | 0.06 | 2.03 | 0.21 | 1.81 | 0.06 | 2.03 | 0.21 |
| 553 | 1.81 | 0.05 | 2.03 | 0.20 | 1.81 | 0.06 | 2.03 | 0.21 | 1.80 | 0.06 | 2.03 | 0.21 |
| 554 | 1.80 | 0.05 | 2.03 | 0.20 | 1.80 | 0.06 | 2.02 | 0.20 | 1.80 | 0.06 | 2.02 | 0.20 |
| 555 | 1.80 | 0.05 | 2.02 | 0.19 | 1.80 | 0.06 | 2.02 | 0.20 | 1.80 | 0.06 | 2.02 | 0.20 |
| 556 | 1.80 | 0.05 | 2.02 | 0.19 | 1.80 | 0.06 | 2.02 | 0.20 | 1.79 | 0.06 | 2.02 | 0.20 |
| 557 | 1.80 | 0.05 | 2.02 | 0.19 | 1.79 | 0.06 | 2.01 | 0.19 | 1.79 | 0.06 | 2.01 | 0.19 |
| 558 | 1.79 | 0.05 | 2.01 | 0.18 | 1.79 | 0.06 | 2.01 | 0.19 | 1.79 | 0.06 | 2.01 | 0.19 |
| 559 | 1.79 | 0.05 | 2.01 | 0.18 | 1.79 | 0.06 | 2.01 | 0.19 | 1.79 | 0.06 | 2.01 | 0.19 |
| 560 | 1.79 | 0.05 | 2.01 | 0.18 | 1.79 | 0.06 | 2.00 | 0.18 | 1.78 | 0.06 | 2.00 | 0.18 |
| 561 | 1.79 | 0.05 | 2.00 | 0.18 | 1.78 | 0.06 | 2.00 | 0.18 | 1.78 | 0.06 | 2.00 | 0.18 |
| 562 | 1.78 | 0.05 | 2.00 | 0.17 | 1.78 | 0.06 | 2.00 | 0.18 | 1.78 | 0.06 | 1.99 | 0.18 |
| 563 | 1.78 | 0.05 | 1.99 | 0.17 | 1.78 | 0.06 | 1.99 | 0.18 | 1.77 | 0.06 | 1.99 | 0.18 |
| 564 | 1.78 | 0.05 | 1.99 | 0.17 | 1.78 | 0.06 | 1.99 | 0.17 | 1.77 | 0.06 | 1.99 | 0.17 |
| 565 | 1.78 | 0.05 | 1.99 | 0.17 | 1.77 | 0.06 | 1.99 | 0.17 | 1.77 | 0.06 | 1.98 | 0.17 |
| 566 | 1.77 | 0.05 | 1.98 | 0.16 | 1.77 | 0.06 | 1.98 | 0.17 | 1.77 | 0.06 | 1.98 | 0.17 |
| 567 | 1.77 | 0.05 | 1.98 | 0.16 | 1.77 | 0.06 | 1.98 | 0.17 | 1.76 | 0.06 | 1.98 | 0.17 |
| 568 | 1.77 | 0.05 | 1.98 | 0.16 | 1.77 | 0.06 | 1.97 | 0.16 | 1.76 | 0.06 | 1.97 | 0.16 |
| 569 | 1.77 | 0.05 | 1.97 | 0.16 | 1.76 | 0.06 | 1.97 | 0.16 | 1.76 | 0.06 | 1.97 | 0.16 |
| 570 | 1.77 | 0.05 | 1.97 | 0.16 | 1.76 | 0.06 | 1.97 | 0.16 | 1.76 | 0.06 | 1.96 | 0.16 |
| 571 | 1.76 | 0.05 | 1.97 | 0.15 | 1.76 | 0.06 | 1.96 | 0.16 | 1.75 | 0.06 | 1.96 | 0.16 |
| 572 | 1.76 | 0.06 | 1.96 | 0.15 | 1.76 | 0.06 | 1.96 | 0.16 | 1.75 | 0.07 | 1.96 | 0.16 |
| 573 | 1.76 | 0.06 | 1.96 | 0.15 | 1.75 | 0.06 | 1.96 | 0.16 | 1.75 | 0.07 | 1.95 | 0.16 |
| 574 | 1.76 | 0.06 | 1.95 | 0.15 | 1.75 | 0.06 | 1.95 | 0.15 | 1.75 | 0.07 | 1.95 | 0.15 |
| 575 | 1.75 | 0.06 | 1.95 | 0.15 | 1.75 | 0.07 | 1.95 | 0.15 | 1.74 | 0.07 | 1.94 | 0.15 |
| 576 | 1.75 | 0.06 | 1.95 | 0.15 | 1.75 | 0.07 | 1.94 | 0.15 | 1.74 | 0.07 | 1.94 | 0.15 |
| 577 | 1.75 | 0.06 | 1.94 | 0.14 | 1.74 | 0.07 | 1.94 | 0.15 | 1.74 | 0.07 | 1.94 | 0.15 |
| 578 | 1.75 | 0.06 | 1.94 | 0.14 | 1.74 | 0.07 | 1.94 | 0.15 | 1.74 | 0.07 | 1.93 | 0.15 |
| 579 | 1.75 | 0.06 | 1.94 | 0.14 | 1.74 | 0.07 | 1.93 | 0.15 | 1.73 | 0.07 | 1.93 | 0.15 |
| 580 | 1.74 | 0.06 | 1.93 | 0.14 | 1.74 | 0.07 | 1.93 | 0.15 | 1.73 | 0.08 | 1.93 | 0.15 |
| 581 | 1.74 | 0.07 | 1.93 | 0.14 | 1.73 | 0.07 | 1.93 | 0.15 | 1.73 | 0.08 | 1.92 | 0.15 |
| 582 | 1.74 | 0.07 | 1.93 | 0.14 | 1.73 | 0.08 | 1.92 | 0.14 | 1.73 | 0.08 | 1.92 | 0.15 |
| 583 | 1.74 | 0.07 | 1.92 | 0.14 | 1.73 | 0.08 | 1.92 | 0.14 | 1.73 | 0.08 | 1.92 | 0.14 |
| 584 | 1.73 | 0.07 | 1.92 | 0.14 | 1.73 | 0.08 | 1.92 | 0.14 | 1.72 | 0.08 | 1.91 | 0.14 |
| 585 | 1.73 | 0.07 | 1.91 | 0.14 | 1.73 | 0.08 | 1.91 | 0.14 | 1.72 | 0.09 | 1.91 | 0.14 |
| 586 | 1.73 | 0.07 | 1.91 | 0.14 | 1.72 | 0.08 | 1.91 | 0.14 | 1.72 | 0.09 | 1.90 | 0.14 |
| 587 | 1.73 | 0.07 | 1.91 | 0.14 | 1.72 | 0.09 | 1.90 | 0.14 | 1.72 | 0.09 | 1.90 | 0.14 |
| 588 | 1.73 | 0.08 | 1.90 | 0.13 | 1.72 | 0.09 | 1.90 | 0.14 | 1.72 | 0.09 | 1.90 | 0.14 |
| 589 | 1.72 | 0.08 | 1.90 | 0.13 | 1.72 | 0.09 | 1.90 | 0.14 | 1.71 | 0.10 | 1.89 | 0.14 |
| 590 | 1.72 | 0.08 | 1.90 | 0.13 | 1.72 | 0.09 | 1.89 | 0.14 | 1.71 | 0.10 | 1.89 | 0.14 |
| 591 | 1.72 | 0.08 | 1.89 | 0.13 | 1.71 | 0.10 | 1.89 | 0.14 | 1.71 | 0.10 | 1.89 | 0.14 |
| 592 | 1.72 | 0.09 | 1.89 | 0.13 | 1.71 | 0.10 | 1.89 | 0.14 | 1.71 | 0.10 | 1.88 | 0.14 |
| 593 | 1.72 | 0.09 | 1.89 | 0.13 | 1.71 | 0.10 | 1.88 | 0.14 | 1.71 | 0.11 | 1.88 | 0.14 |
| 594 | 1.72 | 0.09 | 1.88 | 0.13 | 1.71 | 0.10 | 1.88 | 0.14 | 1.71 | 0.11 | 1.88 | 0.14 |
| 595 | 1.71 | 0.09 | 1.88 | 0.13 | 1.71 | 0.11 | 1.88 | 0.14 | 1.71 | 0.11 | 1.87 | 0.14 |
| 596 | 1.71 | 0.10 | 1.88 | 0.13 | 1.71 | 0.11 | 1.87 | 0.14 | 1.71 | 0.12 | 1.87 | 0.14 |
| 597 | 1.71 | 0.10 | 1.87 | 0.13 | 1.71 | 0.11 | 1.87 | 0.14 | 1.71 | 0.12 | 1.87 | 0.14 |
| 598 | 1.71 | 0.10 | 1.87 | 0.13 | 1.71 | 0.12 | 1.87 | 0.14 | 1.71 | 0.13 | 1.86 | 0.14 |
| 599 | 1.71 | 0.11 | 1.87 | 0.13 | 1.71 | 0.12 | 1.86 | 0.14 | 1.71 | 0.13 | 1.86 | 0.14 |
| 600 | 1.71 | 0.11 | 1.86 | 0.13 | 1.71 | 0.13 | 1.86 | 0.14 | 1.71 | 0.13 | 1.86 | 0.14 |
| 601 | 1.71 | 0.11 | 1.86 | 0.13 | 1.71 | 0.13 | 1.86 | 0.14 | 1.71 | 0.13 | 1.85 | 0.14 |
| 602 | 1.71 | 0.12 | 1.86 | 0.13 | 1.71 | 0.13 | 1.85 | 0.14 | 1.71 | 0.14 | 1.85 | 0.14 |
| 603 | 1.71 | 0.12 | 1.85 | 0.13 | 1.71 | 0.13 | 1.85 | 0.14 | 1.71 | 0.14 | 1.85 | 0.14 |
| 604 | 1.71 | 0.12 | 1.85 | 0.13 | 1.72 | 0.14 | 1.85 | 0.14 | 1.71 | 0.14 | 1.84 | 0.14 |
| 605 | 1.72 | 0.13 | 1.85 | 0.13 | 1.72 | 0.14 | 1.84 | 0.14 | 1.72 | 0.14 | 1.84 | 0.14 |
| 606 | 1.72 | 0.13 | 1.85 | 0.13 | 1.72 | 0.14 | 1.84 | 0.14 | 1.72 | 0.15 | 1.84 | 0.14 |
| 607 | 1.72 | 0.13 | 1.84 | 0.13 | 1.72 | 0.14 | 1.84 | 0.14 | 1.72 | 0.15 | 1.83 | 0.14 |
| 608 | 1.72 | 0.13 | 1.84 | 0.13 | 1.72 | 0.14 | 1.83 | 0.14 | 1.72 | 0.15 | 1.83 | 0.14 |
| 609 | 1.72 | 0.13 | 1.84 | 0.13 | 1.72 | 0.14 | 1.83 | 0.14 | 1.72 | 0.15 | 1.83 | 0.14 |
| 610 | 1.73 | 0.14 | 1.83 | 0.13 | 1.73 | 0.14 | 1.83 | 0.14 | 1.72 | 0.15 | 1.82 | 0.14 |
| 611 | 1.73 | 0.14 | 1.83 | 0.13 | 1.73 | 0.14 | 1.82 | 0.14 | 1.73 | 0.15 | 1.82 | 0.14 |
| 612 | 1.73 | 0.14 | 1.83 | 0.13 | 1.73 | 0.14 | 1.82 | 0.14 | 1.73 | 0.15 | 1.82 | 0.14 |
| 613 | 1.73 | 0.14 | 1.82 | 0.13 | 1.73 | 0.14 | 1.82 | 0.14 | 1.73 | 0.15 | 1.81 | 0.14 |
| 614 | 1.73 | 0.14 | 1.82 | 0.13 | 1.73 | 0.14 | 1.82 | 0.14 | 1.73 | 0.15 | 1.81 | 0.14 |
| 615 | 1.73 | 0.14 | 1.82 | 0.13 | 1.73 | 0.14 | 1.81 | 0.14 | 1.73 | 0.15 | 1.81 | 0.14 |
| 616 | 1.73 | 0.14 | 1.81 | 0.13 | 1.73 | 0.14 | 1.81 | 0.14 | 1.73 | 0.15 | 1.80 | 0.14 |
| 617 | 1.73 | 0.14 | 1.81 | 0.13 | 1.73 | 0.14 | 1.81 | 0.14 | 1.73 | 0.15 | 1.80 | 0.14 |
| 618 | 1.73 | 0.13 | 1.81 | 0.13 | 1.73 | 0.14 | 1.80 | 0.14 | 1.73 | 0.15 | 1.80 | 0.14 |
| 619 | 1.73 | 0.13 | 1.81 | 0.13 | 1.73 | 0.14 | 1.80 | 0.14 | 1.73 | 0.15 | 1.80 | 0.14 |
| 620 | 1.73 | 0.13 | 1.80 | 0.13 | 1.73 | 0.14 | 1.80 | 0.14 | 1.73 | 0.15 | 1.79 | 0.14 |
| 621 | 1.73 | 0.14 | 1.80 | 0.13 | 1.73 | 0.14 | 1.79 | 0.14 | 1.73 | 0.15 | 1.79 | 0.14 |
| 622 | 1.73 | 0.14 | 1.80 | 0.13 | 1.73 | 0.15 | 1.79 | 0.14 | 1.73 | 0.15 | 1.79 | 0.14 |
| 623 | 1.73 | 0.14 | 1.79 | 0.13 | 1.73 | 0.15 | 1.79 | 0.14 | 1.73 | 0.15 | 1.78 | 0.14 |
| 624 | 1.73 | 0.14 | 1.79 | 0.13 | 1.73 | 0.15 | 1.78 | 0.14 | 1.73 | 0.16 | 1.78 | 0.14 |
| 625 | 1.73 | 0.14 | 1.79 | 0.14 | 1.73 | 0.15 | 1.78 | 0.14 | 1.73 | 0.16 | 1.78 | 0.14 |
| 626 | 1.73 | 0.14 | 1.78 | 0.14 | 1.73 | 0.15 | 1.78 | 0.14 | 1.73 | 0.16 | 1.77 | 0.15 |
| 627 | 1.73 | 0.14 | 1.78 | 0.14 | 1.73 | 0.15 | 1.78 | 0.14 | 1.73 | 0.16 | 1.77 | 0.15 |
| 628 | 1.73 | 0.14 | 1.78 | 0.14 | 1.73 | 0.15 | 1.77 | 0.14 | 1.73 | 0.16 | 1.77 | 0.15 |
| 629 | 1.73 | 0.14 | 1.78 | 0.14 | 1.73 | 0.16 | 1.77 | 0.15 | 1.73 | 0.16 | 1.77 | 0.15 |
| 630 | 1.73 | 0.14 | 1.77 | 0.14 | 1.73 | 0.16 | 1.77 | 0.15 | 1.73 | 0.16 | 1.76 | 0.15 |
| 631 | 1.73 | 0.15 | 1.77 | 0.14 | 1.73 | 0.16 | 1.76 | 0.15 | 1.73 | 0.17 | 1.76 | 0.15 |
| 632 | 1.73 | 0.15 | 1.77 | 0.14 | 1.73 | 0.16 | 1.76 | 0.15 | 1.73 | 0.17 | 1.76 | 0.15 |
| 633 | 1.73 | 0.15 | 1.76 | 0.14 | 1.73 | 0.16 | 1.76 | 0.15 | 1.73 | 0.17 | 1.75 | 0.15 |
| 634 | 1.74 | 0.15 | 1.76 | 0.14 | 1.73 | 0.16 | 1.75 | 0.15 | 1.74 | 0.17 | 1.75 | 0.15 |
| 635 | 1.74 | 0.15 | 1.76 | 0.14 | 1.74 | 0.16 | 1.75 | 0.15 | 1.74 | 0.17 | 1.75 | 0.15 |
| 636 | 1.74 | 0.15 | 1.75 | 0.14 | 1.74 | 0.17 | 1.75 | 0.15 | 1.74 | 0.17 | 1.74 | 0.15 |
| 637 | 1.74 | 0.15 | 1.75 | 0.14 | 1.74 | 0.17 | 1.75 | 0.15 | 1.74 | 0.17 | 1.74 | 0.16 |
| 638 | 1.74 | 0.15 | 1.75 | 0.15 | 1.74 | 0.17 | 1.74 | 0.15 | 1.74 | 0.17 | 1.74 | 0.16 |
| 639 | 1.74 | 0.15 | 1.75 | 0.15 | 1.74 | 0.17 | 1.74 | 0.15 | 1.75 | 0.17 | 1.74 | 0.16 |
| 640 | 1.74 | 0.15 | 1.74 | 0.15 | 1.74 | 0.17 | 1.74 | 0.16 | 1.75 | 0.17 | 1.73 | 0.16 |
| 641 | 1.74 | 0.16 | 1.74 | 0.15 | 1.75 | 0.17 | 1.73 | 0.16 | 1.75 | 0.17 | 1.73 | 0.16 |
| 642 | 1.75 | 0.16 | 1.74 | 0.15 | 1.75 | 0.17 | 1.73 | 0.16 | 1.75 | 0.17 | 1.73 | 0.16 |
| 643 | 1.75 | 0.16 | 1.73 | 0.15 | 1.75 | 0.17 | 1.73 | 0.16 | 1.75 | 0.17 | 1.72 | 0.16 |
| 644 | 1.75 | 0.16 | 1.73 | 0.15 | 1.75 | 0.16 | 1.73 | 0.16 | 1.75 | 0.17 | 1.72 | 0.16 |
| 645 | 1.75 | 0.15 | 1.73 | 0.15 | 1.75 | 0.16 | 1.72 | 0.16 | 1.75 | 0.17 | 1.72 | 0.16 |
| 646 | 1.75 | 0.15 | 1.73 | 0.15 | 1.75 | 0.16 | 1.72 | 0.16 | 1.76 | 0.16 | 1.71 | 0.17 |
| 647 | 1.75 | 0.15 | 1.72 | 0.16 | 1.75 | 0.16 | 1.72 | 0.16 | 1.76 | 0.16 | 1.71 | 0.17 |
| 648 | 1.76 | 0.15 | 1.72 | 0.16 | 1.76 | 0.16 | 1.71 | 0.17 | 1.76 | 0.16 | 1.71 | 0.17 |
| 649 | 1.76 | 0.15 | 1.72 | 0.16 | 1.76 | 0.16 | 1.71 | 0.17 | 1.76 | 0.16 | 1.71 | 0.17 |
| 650 | 1.76 | 0.15 | 1.71 | 0.16 | 1.76 | 0.16 | 1.71 | 0.17 | 1.76 | 0.16 | 1.70 | 0.17 |
| 651 | 1.76 | 0.15 | 1.71 | 0.16 | 1.76 | 0.16 | 1.71 | 0.17 | 1.76 | 0.16 | 1.70 | 0.17 |
| 652 | 1.76 | 0.15 | 1.71 | 0.16 | 1.76 | 0.15 | 1.70 | 0.17 | 1.76 | 0.16 | 1.70 | 0.17 |
| 653 | 1.76 | 0.15 | 1.70 | 0.16 | 1.76 | 0.15 | 1.70 | 0.17 | 1.76 | 0.15 | 1.69 | 0.18 |
| 654 | 1.76 | 0.15 | 1.70 | 0.17 | 1.76 | 0.15 | 1.70 | 0.17 | 1.76 | 0.15 | 1.69 | 0.18 |
| 655 | 1.76 | 0.15 | 1.70 | 0.17 | 1.76 | 0.15 | 1.69 | 0.18 | 1.76 | 0.15 | 1.69 | 0.18 |
| 656 | 1.76 | 0.14 | 1.70 | 0.17 | 1.76 | 0.15 | 1.69 | 0.18 | 1.76 | 0.15 | 1.69 | 0.18 |
| 657 | 1.76 | 0.14 | 1.69 | 0.17 | 1.76 | 0.15 | 1.69 | 0.18 | 1.76 | 0.15 | 1.68 | 0.18 |
| 658 | 1.76 | 0.14 | 1.69 | 0.17 | 1.76 | 0.15 | 1.69 | 0.18 | 1.75 | 0.15 | 1.68 | 0.19 |
| 659 | 1.76 | 0.14 | 1.69 | 0.17 | 1.76 | 0.15 | 1.68 | 0.18 | 1.75 | 0.15 | 1.68 | 0.19 |
| 660 | 1.76 | 0.14 | 1.68 | 0.17 | 1.75 | 0.15 | 1.68 | 0.18 | 1.75 | 0.15 | 1.67 | 0.19 |
| 661 | 1.76 | 0.14 | 1.68 | 0.18 | 1.75 | 0.15 | 1.68 | 0.19 | 1.75 | 0.15 | 1.67 | 0.19 |
| 662 | 1.76 | 0.14 | 1.68 | 0.18 | 1.75 | 0.14 | 1.67 | 0.19 | 1.75 | 0.15 | 1.67 | 0.19 |
| 663 | 1.76 | 0.14 | 1.68 | 0.18 | 1.75 | 0.14 | 1.67 | 0.19 | 1.75 | 0.15 | 1.67 | 0.20 |
| 664 | 1.76 | 0.14 | 1.67 | 0.18 | 1.75 | 0.14 | 1.67 | 0.19 | 1.75 | 0.15 | 1.66 | 0.20 |
| 665 | 1.76 | 0.13 | 1.67 | 0.18 | 1.75 | 0.14 | 1.67 | 0.19 | 1.75 | 0.15 | 1.66 | 0.20 |
| 666 | 1.76 | 0.13 | 1.67 | 0.19 | 1.75 | 0.14 | 1.66 | 0.20 | 1.75 | 0.14 | 1.66 | 0.20 |
| 667 | 1.76 | 0.13 | 1.66 | 0.19 | 1.75 | 0.14 | 1.66 | 0.20 | 1.75 | 0.14 | 1.66 | 0.20 |
| 668 | 1.75 | 0.13 | 1.66 | 0.19 | 1.75 | 0.14 | 1.66 | 0.20 | 1.75 | 0.14 | 1.65 | 0.21 |
| 669 | 1.75 | 0.13 | 1.66 | 0.19 | 1.75 | 0.14 | 1.65 | 0.20 | 1.74 | 0.14 | 1.65 | 0.21 |
| 670 | 1.75 | 0.13 | 1.66 | 0.20 | 1.75 | 0.14 | 1.65 | 0.21 | 1.74 | 0.14 | 1.65 | 0.21 |
| 671 | 1.75 | 0.13 | 1.65 | 0.20 | 1.74 | 0.14 | 1.65 | 0.21 | 1.74 | 0.14 | 1.64 | 0.21 |
| 672 | 1.75 | 0.13 | 1.65 | 0.20 | 1.74 | 0.14 | 1.65 | 0.21 | 1.74 | 0.14 | 1.64 | 0.22 |
| 673 | 1.75 | 0.13 | 1.65 | 0.20 | 1.74 | 0.14 | 1.64 | 0.21 | 1.74 | 0.14 | 1.64 | 0.22 |
| 674 | 1.75 | 0.13 | 1.64 | 0.21 | 1.74 | 0.14 | 1.64 | 0.22 | 1.74 | 0.14 | 1.64 | 0.22 |
| 675 | 1.75 | 0.13 | 1.64 | 0.21 | 1.74 | 0.14 | 1.64 | 0.22 | 1.74 | 0.14 | 1.63 | 0.22 |
| 676 | 1.75 | 0.13 | 1.64 | 0.21 | 1.74 | 0.14 | 1.64 | 0.22 | 1.74 | 0.14 | 1.63 | 0.23 |
| 677 | 1.74 | 0.13 | 1.64 | 0.21 | 1.74 | 0.14 | 1.63 | 0.22 | 1.74 | 0.15 | 1.63 | 0.23 |
| 678 | 1.74 | 0.13 | 1.63 | 0.22 | 1.74 | 0.14 | 1.63 | 0.23 | 1.74 | 0.15 | 1.63 | 0.23 |
| 679 | 1.74 | 0.13 | 1.63 | 0.22 | 1.74 | 0.14 | 1.63 | 0.23 | 1.74 | 0.15 | 1.63 | 0.24 |
| 680 | 1.74 | 0.13 | 1.63 | 0.22 | 1.74 | 0.14 | 1.63 | 0.23 | 1.73 | 0.15 | 1.62 | 0.24 |
| 681 | 1.74 | 0.13 | 1.63 | 0.22 | 1.74 | 0.14 | 1.62 | 0.24 | 1.73 | 0.15 | 1.62 | 0.24 |
| 682 | 1.74 | 0.13 | 1.62 | 0.23 | 1.73 | 0.14 | 1.62 | 0.24 | 1.73 | 0.15 | 1.62 | 0.25 |
| 683 | 1.74 | 0.13 | 1.62 | 0.23 | 1.73 | 0.14 | 1.62 | 0.24 | 1.73 | 0.15 | 1.62 | 0.25 |
| 684 | 1.74 | 0.13 | 1.62 | 0.23 | 1.73 | 0.14 | 1.62 | 0.25 | 1.73 | 0.15 | 1.61 | 0.25 |
| 685 | 1.74 | 0.13 | 1.62 | 0.24 | 1.73 | 0.15 | 1.62 | 0.25 | 1.73 | 0.15 | 1.61 | 0.26 |
| 686 | 1.74 | 0.13 | 1.61 | 0.24 | 1.73 | 0.15 | 1.61 | 0.25 | 1.73 | 0.15 | 1.61 | 0.26 |
| 687 | 1.73 | 0.13 | 1.61 | 0.24 | 1.73 | 0.15 | 1.61 | 0.26 | 1.73 | 0.15 | 1.61 | 0.26 |
| 688 | 1.73 | 0.13 | 1.61 | 0.25 | 1.73 | 0.15 | 1.61 | 0.26 | 1.73 | 0.15 | 1.61 | 0.27 |
| 689 | 1.73 | 0.13 | 1.61 | 0.25 | 1.73 | 0.15 | 1.61 | 0.26 | 1.73 | 0.15 | 1.60 | 0.27 |
| 690 | 1.73 | 0.13 | 1.61 | 0.26 | 1.73 | 0.15 | 1.61 | 0.27 | 1.73 | 0.15 | 1.60 | 0.27 |
| 691 | 1.73 | 0.13 | 1.60 | 0.26 | 1.73 | 0.15 | 1.60 | 0.27 | 1.73 | 0.15 | 1.60 | 0.28 |
| 692 | 1.73 | 0.13 | 1.60 | 0.26 | 1.73 | 0.15 | 1.60 | 0.27 | 1.73 | 0.15 | 1.60 | 0.28 |
| 693 | 1.73 | 0.13 | 1.60 | 0.27 | 1.73 | 0.15 | 1.60 | 0.28 | 1.72 | 0.15 | 1.60 | 0.28 |
| 694 | 1.73 | 0.13 | 1.60 | 0.27 | 1.73 | 0.15 | 1.60 | 0.28 | 1.72 | 0.15 | 1.60 | 0.29 |
| 695 | 1.73 | 0.14 | 1.60 | 0.27 | 1.73 | 0.15 | 1.60 | 0.28 | 1.72 | 0.16 | 1.60 | 0.29 |
| 696 | 1.73 | 0.14 | 1.59 | 0.28 | 1.72 | 0.15 | 1.60 | 0.29 | 1.72 | 0.16 | 1.59 | 0.30 |
| 697 | 1.72 | 0.14 | 1.59 | 0.28 | 1.72 | 0.15 | 1.60 | 0.29 | 1.72 | 0.16 | 1.59 | 0.30 |
| 698 | 1.72 | 0.14 | 1.59 | 0.29 | 1.72 | 0.15 | 1.59 | 0.30 | 1.72 | 0.16 | 1.59 | 0.30 |
| 699 | 1.72 | 0.14 | 1.59 | 0.29 | 1.72 | 0.15 | 1.59 | 0.30 | 1.72 | 0.16 | 1.59 | 0.31 |
| 700 | 1.72 | 0.14 | 1.59 | 0.30 | 1.72 | 0.16 | 1.59 | 0.30 | 1.72 | 0.16 | 1.59 | 0.31 |
| 701 | 1.72 | 0.14 | 1.59 | 0.30 | 1.72 | 0.16 | 1.59 | 0.31 | 1.72 | 0.16 | 1.59 | 0.32 |
| 702 | 1.72 | 0.14 | 1.59 | 0.30 | 1.72 | 0.16 | 1.59 | 0.31 | 1.72 | 0.16 | 1.59 | 0.32 |
| 703 | 1.72 | 0.14 | 1.59 | 0.31 | 1.72 | 0.16 | 1.59 | 0.32 | 1.72 | 0.16 | 1.59 | 0.32 |
| 704 | 1.72 | 0.14 | 1.59 | 0.31 | 1.72 | 0.16 | 1.59 | 0.32 | 1.72 | 0.16 | 1.59 | 0.33 |
| 705 | 1.72 | 0.14 | 1.58 | 0.32 | 1.72 | 0.16 | 1.59 | 0.32 | 1.72 | 0.16 | 1.59 | 0.33 |
| 706 | 1.72 | 0.14 | 1.58 | 0.32 | 1.72 | 0.16 | 1.59 | 0.33 | 1.72 | 0.16 | 1.59 | 0.34 |
| 707 | 1.72 | 0.15 | 1.58 | 0.33 | 1.72 | 0.16 | 1.59 | 0.33 | 1.72 | 0.16 | 1.59 | 0.34 |
| 708 | 1.72 | 0.15 | 1.58 | 0.33 | 1.72 | 0.16 | 1.59 | 0.34 | 1.72 | 0.17 | 1.59 | 0.35 |
| 709 | 1.72 | 0.15 | 1.58 | 0.34 | 1.72 | 0.16 | 1.59 | 0.34 | 1.72 | 0.17 | 1.59 | 0.35 |
| 710 | 1.72 | 0.15 | 1.58 | 0.34 | 1.72 | 0.16 | 1.59 | 0.35 | 1.72 | 0.17 | 1.59 | 0.35 |
| 711 | 1.72 | 0.15 | 1.58 | 0.35 | 1.72 | 0.16 | 1.59 | 0.35 | 1.72 | 0.17 | 1.59 | 0.36 |
| 712 | 1.72 | 0.15 | 1.59 | 0.35 | 1.72 | 0.16 | 1.59 | 0.35 | 1.72 | 0.17 | 1.59 | 0.36 |
| 713 | 1.71 | 0.15 | 1.59 | 0.36 | 1.72 | 0.17 | 1.59 | 0.36 | 1.72 | 0.17 | 1.59 | 0.37 |
| 714 | 1.71 | 0.15 | 1.59 | 0.36 | 1.72 | 0.17 | 1.59 | 0.36 | 1.72 | 0.17 | 1.60 | 0.37 |
| 715 | 1.71 | 0.15 | 1.59 | 0.36 | 1.72 | 0.17 | 1.59 | 0.37 | 1.72 | 0.17 | 1.60 | 0.37 |
| 716 | 1.71 | 0.15 | 1.59 | 0.37 | 1.72 | 0.17 | 1.60 | 0.37 | 1.72 | 0.17 | 1.60 | 0.38 |
| 717 | 1.71 | 0.16 | 1.59 | 0.37 | 1.72 | 0.17 | 1.60 | 0.37 | 1.72 | 0.17 | 1.60 | 0.38 |
| 718 | 1.71 | 0.16 | 1.59 | 0.38 | 1.72 | 0.17 | 1.60 | 0.38 | 1.72 | 0.17 | 1.60 | 0.38 |
| 719 | 1.71 | 0.16 | 1.59 | 0.38 | 1.72 | 0.17 | 1.60 | 0.38 | 1.72 | 0.17 | 1.60 | 0.39 |
| 720 | 1.71 | 0.16 | 1.60 | 0.39 | 1.72 | 0.17 | 1.60 | 0.38 | 1.72 | 0.17 | 1.61 | 0.39 |
| 721 | 1.71 | 0.16 | 1.60 | 0.39 | 1.72 | 0.17 | 1.60 | 0.39 | 1.72 | 0.18 | 1.61 | 0.39 |
| 722 | 1.71 | 0.16 | 1.60 | 0.39 | 1.72 | 0.17 | 1.61 | 0.39 | 1.72 | 0.18 | 1.61 | 0.40 |
| 723 | 1.71 | 0.16 | 1.60 | 0.40 | 1.72 | 0.17 | 1.61 | 0.39 | 1.72 | 0.18 | 1.61 | 0.40 |
| 724 | 1.71 | 0.16 | 1.60 | 0.40 | 1.72 | 0.17 | 1.61 | 0.40 | 1.72 | 0.18 | 1.61 | 0.40 |
| 725 | 1.71 | 0.16 | 1.61 | 0.40 | 1.72 | 0.17 | 1.61 | 0.40 | 1.72 | 0.18 | 1.62 | 0.40 |
| 726 | 1.71 | 0.17 | 1.61 | 0.41 | 1.72 | 0.18 | 1.61 | 0.40 | 1.72 | 0.18 | 1.62 | 0.41 |
| 727 | 1.71 | 0.17 | 1.61 | 0.41 | 1.72 | 0.18 | 1.62 | 0.40 | 1.72 | 0.18 | 1.62 | 0.41 |
| 728 | 1.71 | 0.17 | 1.62 | 0.41 | 1.72 | 0.18 | 1.62 | 0.41 | 1.72 | 0.18 | 1.62 | 0.41 |
| 729 | 1.71 | 0.17 | 1.62 | 0.41 | 1.72 | 0.18 | 1.62 | 0.41 | 1.72 | 0.18 | 1.62 | 0.41 |
| 730 | 1.71 | 0.17 | 1.62 | 0.42 | 1.72 | 0.18 | 1.62 | 0.41 | 1.72 | 0.18 | 1.63 | 0.41 |
| 731 | 1.71 | 0.17 | 1.62 | 0.42 | 1.72 | 0.18 | 1.62 | 0.41 | 1.72 | 0.18 | 1.63 | 0.42 |
| 732 | 1.71 | 0.17 | 1.63 | 0.42 | 1.72 | 0.18 | 1.63 | 0.41 | 1.72 | 0.18 | 1.63 | 0.42 |
| 733 | 1.71 | 0.17 | 1.63 | 0.42 | 1.72 | 0.18 | 1.63 | 0.42 | 1.72 | 0.18 | 1.63 | 0.42 |
| 734 | 1.71 | 0.17 | 1.63 | 0.42 | 1.72 | 0.18 | 1.63 | 0.42 | 1.72 | 0.18 | 1.64 | 0.42 |
| 735 | 1.71 | 0.17 | 1.64 | 0.43 | 1.72 | 0.18 | 1.63 | 0.42 | 1.72 | 0.18 | 1.64 | 0.42 |
| 736 | 1.71 | 0.18 | 1.64 | 0.43 | 1.72 | 0.18 | 1.64 | 0.42 | 1.72 | 0.19 | 1.64 | 0.42 |
| 737 | 1.71 | 0.18 | 1.64 | 0.43 | 1.72 | 0.18 | 1.64 | 0.42 | 1.72 | 0.19 | 1.64 | 0.42 |
| 738 | 1.71 | 0.18 | 1.64 | 0.43 | 1.72 | 0.18 | 1.64 | 0.42 | 1.72 | 0.19 | 1.64 | 0.42 |
| 739 | 1.71 | 0.18 | 1.65 | 0.43 | 1.72 | 0.18 | 1.64 | 0.42 | 1.72 | 0.19 | 1.65 | 0.42 |
| 740 | 1.71 | 0.18 | 1.65 | 0.43 | 1.72 | 0.18 | 1.64 | 0.42 | 1.72 | 0.19 | 1.65 | 0.42 |
| 741 | 1.71 | 0.18 | 1.65 | 0.43 | 1.72 | 0.18 | 1.65 | 0.42 | 1.72 | 0.19 | 1.65 | 0.42 |
| 742 | 1.71 | 0.18 | 1.65 | 0.43 | 1.72 | 0.19 | 1.65 | 0.42 | 1.72 | 0.19 | 1.65 | 0.42 |
| 743 | 1.72 | 0.18 | 1.66 | 0.43 | 1.72 | 0.19 | 1.65 | 0.42 | 1.72 | 0.19 | 1.65 | 0.42 |
| 744 | 1.72 | 0.18 | 1.66 | 0.43 | 1.72 | 0.19 | 1.65 | 0.42 | 1.72 | 0.19 | 1.65 | 0.42 |
| 745 | 1.72 | 0.18 | 1.66 | 0.43 | 1.72 | 0.19 | 1.65 | 0.43 | 1.72 | 0.19 | 1.66 | 0.42 |
| 746 | 1.72 | 0.18 | 1.66 | 0.43 | 1.72 | 0.19 | 1.65 | 0.43 | 1.72 | 0.19 | 1.66 | 0.42 |
| 747 | 1.72 | 0.19 | 1.66 | 0.43 | 1.72 | 0.19 | 1.66 | 0.43 | 1.72 | 0.19 | 1.66 | 0.42 |
| 748 | 1.72 | 0.19 | 1.67 | 0.43 | 1.72 | 0.19 | 1.66 | 0.43 | 1.72 | 0.19 | 1.66 | 0.42 |
| 749 | 1.72 | 0.19 | 1.67 | 0.43 | 1.72 | 0.19 | 1.66 | 0.43 | 1.72 | 0.19 | 1.66 | 0.42 |
| 750 | 1.72 | 0.19 | 1.67 | 0.43 | 1.72 | 0.19 | 1.66 | 0.43 | 1.72 | 0.19 | 1.66 | 0.42 |
| 751 | 1.72 | 0.19 | 1.67 | 0.43 | 1.72 | 0.19 | 1.66 | 0.43 | 1.72 | 0.19 | 1.66 | 0.42 |
| 752 | 1.72 | 0.19 | 1.67 | 0.43 | 1.72 | 0.19 | 1.66 | 0.42 | 1.72 | 0.19 | 1.66 | 0.42 |
| 753 | 1.72 | 0.19 | 1.67 | 0.43 | 1.72 | 0.19 | 1.66 | 0.42 | 1.72 | 0.19 | 1.66 | 0.42 |
| 754 | 1.72 | 0.19 | 1.67 | 0.43 | 1.72 | 0.19 | 1.66 | 0.42 | 1.72 | 0.19 | 1.66 | 0.42 |
| 755 | 1.72 | 0.19 | 1.67 | 0.43 | 1.72 | 0.19 | 1.66 | 0.42 | 1.72 | 0.20 | 1.66 | 0.42 |
| 756 | 1.72 | 0.19 | 1.67 | 0.43 | 1.72 | 0.19 | 1.66 | 0.42 | 1.72 | 0.20 | 1.66 | 0.42 |
| 757 | 1.72 | 0.19 | 1.67 | 0.43 | 1.72 | 0.19 | 1.66 | 0.42 | 1.72 | 0.20 | 1.66 | 0.42 |
| 758 | 1.72 | 0.19 | 1.67 | 0.43 | 1.72 | 0.19 | 1.66 | 0.42 | 1.72 | 0.20 | 1.66 | 0.42 |
| 759 | 1.72 | 0.19 | 1.67 | 0.43 | 1.72 | 0.19 | 1.66 | 0.42 | 1.72 | 0.20 | 1.66 | 0.42 |
| 760 | 1.72 | 0.19 | 1.67 | 0.42 | 1.72 | 0.19 | 1.66 | 0.42 | 1.72 | 0.20 | 1.66 | 0.42 |
| 761 | 1.72 | 0.19 | 1.67 | 0.42 | 1.72 | 0.19 | 1.66 | 0.42 | 1.72 | 0.20 | 1.66 | 0.42 |
| 762 | 1.72 | 0.19 | 1.67 | 0.42 | 1.72 | 0.20 | 1.66 | 0.42 | 1.72 | 0.20 | 1.66 | 0.42 |
| 763 | 1.72 | 0.19 | 1.67 | 0.42 | 1.72 | 0.20 | 1.66 | 0.42 | 1.72 | 0.20 | 1.66 | 0.42 |
| 764 | 1.72 | 0.19 | 1.67 | 0.42 | 1.72 | 0.20 | 1.66 | 0.42 | 1.72 | 0.20 | 1.66 | 0.42 |
| 765 | 1.72 | 0.19 | 1.67 | 0.42 | 1.72 | 0.20 | 1.66 | 0.42 | 1.72 | 0.20 | 1.66 | 0.42 |
| 766 | 1.72 | 0.19 | 1.67 | 0.42 | 1.72 | 0.20 | 1.66 | 0.42 | 1.72 | 0.20 | 1.66 | 0.42 |
| 767 | 1.73 | 0.19 | 1.67 | 0.42 | 1.72 | 0.20 | 1.66 | 0.42 | 1.72 | 0.20 | 1.66 | 0.42 |
| 768 | 1.73 | 0.19 | 1.67 | 0.42 | 1.72 | 0.20 | 1.66 | 0.42 | 1.72 | 0.20 | 1.66 | 0.42 |
| 769 | 1.73 | 0.19 | 1.67 | 0.42 | 1.72 | 0.20 | 1.66 | 0.42 | 1.72 | 0.20 | 1.66 | 0.42 |
| 770 | 1.73 | 0.20 | 1.67 | 0.42 | 1.72 | 0.20 | 1.66 | 0.42 | 1.72 | 0.20 | 1.66 | 0.42 |
| 771 | 1.73 | 0.20 | 1.67 | 0.42 | 1.72 | 0.20 | 1.66 | 0.42 | 1.72 | 0.20 | 1.66 | 0.42 |
| 772 | 1.73 | 0.20 | 1.67 | 0.42 | 1.72 | 0.20 | 1.66 | 0.43 | 1.72 | 0.20 | 1.66 | 0.42 |
| 773 | 1.73 | 0.20 | 1.67 | 0.42 | 1.72 | 0.20 | 1.66 | 0.43 | 1.72 | 0.20 | 1.66 | 0.42 |
| 774 | 1.73 | 0.20 | 1.67 | 0.42 | 1.72 | 0.20 | 1.66 | 0.43 | 1.72 | 0.20 | 1.65 | 0.43 |
| 775 | 1.73 | 0.20 | 1.67 | 0.43 | 1.72 | 0.20 | 1.66 | 0.43 | 1.72 | 0.20 | 1.65 | 0.43 |
| 776 | 1.73 | 0.20 | 1.66 | 0.43 | 1.72 | 0.20 | 1.65 | 0.43 | 1.72 | 0.20 | 1.65 | 0.43 |
| 777 | 1.73 | 0.20 | 1.66 | 0.43 | 1.72 | 0.20 | 1.65 | 0.43 | 1.72 | 0.20 | 1.65 | 0.43 |
| 778 | 1.73 | 0.20 | 1.66 | 0.43 | 1.73 | 0.20 | 1.65 | 0.43 | 1.72 | 0.20 | 1.65 | 0.43 |
| 779 | 1.73 | 0.20 | 1.66 | 0.43 | 1.73 | 0.20 | 1.65 | 0.43 | 1.72 | 0.20 | 1.65 | 0.43 |
| 780 | 1.73 | 0.20 | 1.66 | 0.43 | 1.73 | 0.20 | 1.65 | 0.43 | 1.72 | 0.20 | 1.65 | 0.43 |
| 781 | 1.73 | 0.20 | 1.66 | 0.43 | 1.73 | 0.20 | 1.65 | 0.43 | 1.72 | 0.21 | 1.65 | 0.43 |
| 782 | 1.73 | 0.20 | 1.66 | 0.43 | 1.73 | 0.20 | 1.65 | 0.43 | 1.72 | 0.21 | 1.65 | 0.43 |
| 783 | 1.73 | 0.20 | 1.66 | 0.43 | 1.73 | 0.20 | 1.65 | 0.43 | 1.72 | 0.21 | 1.65 | 0.43 |
| 784 | 1.73 | 0.20 | 1.66 | 0.43 | 1.73 | 0.20 | 1.65 | 0.43 | 1.72 | 0.21 | 1.64 | 0.43 |
| 785 | 1.73 | 0.20 | 1.65 | 0.43 | 1.73 | 0.20 | 1.65 | 0.44 | 1.72 | 0.21 | 1.64 | 0.44 |
| 786 | 1.73 | 0.20 | 1.65 | 0.44 | 1.73 | 0.20 | 1.65 | 0.44 | 1.73 | 0.21 | 1.64 | 0.44 |
| 787 | 1.73 | 0.20 | 1.65 | 0.44 | 1.73 | 0.20 | 1.64 | 0.44 | 1.73 | 0.21 | 1.64 | 0.44 |
| 788 | 1.73 | 0.20 | 1.65 | 0.44 | 1.73 | 0.20 | 1.64 | 0.44 | 1.73 | 0.21 | 1.64 | 0.44 |
| 789 | 1.73 | 0.20 | 1.65 | 0.44 | 1.73 | 0.21 | 1.64 | 0.44 | 1.73 | 0.21 | 1.64 | 0.44 |
| 790 | 1.73 | 0.20 | 1.65 | 0.44 | 1.73 | 0.21 | 1.64 | 0.44 | 1.73 | 0.21 | 1.64 | 0.44 |
| 791 | 1.73 | 0.20 | 1.65 | 0.44 | 1.73 | 0.21 | 1.64 | 0.44 | 1.73 | 0.21 | 1.64 | 0.44 |
| 792 | 1.73 | 0.20 | 1.65 | 0.44 | 1.73 | 0.21 | 1.64 | 0.45 | 1.73 | 0.21 | 1.64 | 0.45 |
| 793 | 1.73 | 0.20 | 1.65 | 0.45 | 1.73 | 0.21 | 1.64 | 0.45 | 1.73 | 0.21 | 1.64 | 0.45 |
| 794 | 1.73 | 0.20 | 1.65 | 0.45 | 1.73 | 0.21 | 1.64 | 0.45 | 1.73 | 0.21 | 1.63 | 0.45 |
| 795 | 1.73 | 0.20 | 1.65 | 0.45 | 1.73 | 0.21 | 1.64 | 0.45 | 1.73 | 0.21 | 1.63 | 0.45 |
| 796 | 1.73 | 0.20 | 1.64 | 0.45 | 1.73 | 0.21 | 1.64 | 0.45 | 1.73 | 0.21 | 1.63 | 0.45 |
| 797 | 1.73 | 0.20 | 1.64 | 0.45 | 1.73 | 0.21 | 1.64 | 0.45 | 1.73 | 0.21 | 1.63 | 0.45 |
| 798 | 1.73 | 0.20 | 1.64 | 0.46 | 1.73 | 0.21 | 1.63 | 0.46 | 1.73 | 0.21 | 1.63 | 0.46 |
| 799 | 1.73 | 0.20 | 1.64 | 0.46 | 1.73 | 0.21 | 1.63 | 0.46 | 1.73 | 0.21 | 1.63 | 0.46 |
| 800 | 1.73 | 0.20 | 1.64 | 0.46 | 1.73 | 0.21 | 1.63 | 0.46 | 1.73 | 0.21 | 1.63 | 0.46 |
| 801 | 1.73 | 0.20 | 1.64 | 0.46 | 1.73 | 0.21 | 1.63 | 0.46 | 1.73 | 0.21 | 1.63 | 0.46 |
| 802 | 1.73 | 0.20 | 1.64 | 0.46 | 1.73 | 0.21 | 1.63 | 0.46 | 1.73 | 0.21 | 1.63 | 0.46 |
| 803 | 1.73 | 0.20 | 1.64 | 0.47 | 1.73 | 0.21 | 1.63 | 0.46 | 1.73 | 0.21 | 1.63 | 0.46 |
| 804 | 1.73 | 0.20 | 1.64 | 0.47 | 1.73 | 0.21 | 1.63 | 0.47 | 1.73 | 0.21 | 1.63 | 0.47 |
| 805 | 1.73 | 0.20 | 1.64 | 0.47 | 1.73 | 0.21 | 1.63 | 0.47 | 1.73 | 0.21 | 1.63 | 0.47 |
| 806 | 1.73 | 0.20 | 1.64 | 0.47 | 1.73 | 0.21 | 1.63 | 0.47 | 1.73 | 0.21 | 1.63 | 0.47 |
| 807 | 1.73 | 0.21 | 1.64 | 0.47 | 1.73 | 0.21 | 1.63 | 0.47 | 1.73 | 0.21 | 1.62 | 0.47 |
| 808 | 1.73 | 0.21 | 1.64 | 0.48 | 1.73 | 0.21 | 1.63 | 0.48 | 1.73 | 0.21 | 1.62 | 0.47 |
| 809 | 1.73 | 0.21 | 1.64 | 0.48 | 1.73 | 0.21 | 1.63 | 0.48 | 1.73 | 0.22 | 1.62 | 0.48 |
| 810 | 1.73 | 0.21 | 1.64 | 0.48 | 1.73 | 0.21 | 1.63 | 0.48 | 1.73 | 0.22 | 1.62 | 0.48 |
| 811 | 1.73 | 0.21 | 1.64 | 0.48 | 1.73 | 0.21 | 1.63 | 0.48 | 1.73 | 0.22 | 1.62 | 0.48 |
| 812 | 1.73 | 0.21 | 1.64 | 0.49 | 1.73 | 0.21 | 1.63 | 0.48 | 1.73 | 0.22 | 1.62 | 0.48 |
| 813 | 1.73 | 0.21 | 1.64 | 0.49 | 1.73 | 0.21 | 1.63 | 0.49 | 1.73 | 0.22 | 1.62 | 0.49 |
| 814 | 1.73 | 0.21 | 1.64 | 0.49 | 1.73 | 0.21 | 1.63 | 0.49 | 1.73 | 0.22 | 1.62 | 0.49 |
| 815 | 1.73 | 0.21 | 1.64 | 0.49 | 1.73 | 0.21 | 1.63 | 0.49 | 1.73 | 0.22 | 1.62 | 0.49 |
| 816 | 1.73 | 0.21 | 1.64 | 0.49 | 1.73 | 0.22 | 1.63 | 0.49 | 1.73 | 0.22 | 1.62 | 0.49 |
| 817 | 1.73 | 0.21 | 1.64 | 0.50 | 1.73 | 0.22 | 1.62 | 0.49 | 1.73 | 0.22 | 1.62 | 0.49 |
| 818 | 1.73 | 0.21 | 1.64 | 0.50 | 1.73 | 0.22 | 1.62 | 0.50 | 1.73 | 0.22 | 1.62 | 0.50 |
| 819 | 1.73 | 0.21 | 1.64 | 0.50 | 1.73 | 0.22 | 1.62 | 0.50 | 1.73 | 0.22 | 1.62 | 0.50 |
| 820 | 1.73 | 0.21 | 1.64 | 0.50 | 1.73 | 0.22 | 1.62 | 0.50 | 1.73 | 0.22 | 1.62 | 0.50 |
| 821 | 1.73 | 0.21 | 1.64 | 0.51 | 1.73 | 0.22 | 1.62 | 0.50 | 1.73 | 0.22 | 1.62 | 0.50 |
| 822 | 1.73 | 0.21 | 1.64 | 0.51 | 1.73 | 0.22 | 1.62 | 0.51 | 1.73 | 0.22 | 1.62 | 0.51 |
| 823 | 1.73 | 0.21 | 1.64 | 0.51 | 1.73 | 0.22 | 1.62 | 0.51 | 1.73 | 0.22 | 1.62 | 0.51 |
| 824 | 1.73 | 0.21 | 1.64 | 0.51 | 1.73 | 0.22 | 1.62 | 0.51 | 1.73 | 0.22 | 1.62 | 0.51 |
| 825 | 1.73 | 0.21 | 1.64 | 0.52 | 1.73 | 0.22 | 1.62 | 0.51 | 1.73 | 0.22 | 1.62 | 0.51 |
| 826 | 1.73 | 0.21 | 1.64 | 0.52 | 1.73 | 0.22 | 1.62 | 0.51 | 1.73 | 0.22 | 1.62 | 0.51 |
| 827 | 1.73 | 0.21 | 1.64 | 0.52 | 1.73 | 0.22 | 1.62 | 0.52 | 1.73 | 0.22 | 1.62 | 0.52 |
| 828 | 1.73 | 0.21 | 1.64 | 0.52 | 1.73 | 0.22 | 1.62 | 0.52 | 1.73 | 0.22 | 1.62 | 0.52 |
| 829 | 1.73 | 0.21 | 1.64 | 0.52 | 1.73 | 0.22 | 1.62 | 0.52 | 1.73 | 0.22 | 1.62 | 0.52 |
| 830 | 1.73 | 0.21 | 1.64 | 0.53 | 1.73 | 0.22 | 1.62 | 0.52 | 1.73 | 0.22 | 1.62 | 0.52 |
| 831 | 1.73 | 0.21 | 1.64 | 0.53 | 1.73 | 0.22 | 1.62 | 0.53 | 1.73 | 0.22 | 1.62 | 0.53 |
| 832 | 1.73 | 0.21 | 1.64 | 0.53 | 1.73 | 0.22 | 1.63 | 0.53 | 1.73 | 0.22 | 1.62 | 0.53 |
| 833 | 1.73 | 0.22 | 1.64 | 0.53 | 1.73 | 0.22 | 1.63 | 0.53 | 1.73 | 0.22 | 1.62 | 0.53 |
| 834 | 1.73 | 0.22 | 1.64 | 0.54 | 1.73 | 0.22 | 1.63 | 0.53 | 1.73 | 0.22 | 1.62 | 0.53 |
| 835 | 1.73 | 0.22 | 1.64 | 0.54 | 1.73 | 0.22 | 1.63 | 0.54 | 1.73 | 0.23 | 1.62 | 0.53 |
| 836 | 1.73 | 0.22 | 1.64 | 0.54 | 1.73 | 0.22 | 1.63 | 0.54 | 1.73 | 0.23 | 1.62 | 0.54 |
| 837 | 1.73 | 0.22 | 1.64 | 0.54 | 1.73 | 0.22 | 1.63 | 0.54 | 1.73 | 0.23 | 1.62 | 0.54 |
| 838 | 1.73 | 0.22 | 1.64 | 0.54 | 1.73 | 0.22 | 1.63 | 0.54 | 1.73 | 0.23 | 1.62 | 0.54 |
| 839 | 1.73 | 0.22 | 1.64 | 0.55 | 1.73 | 0.22 | 1.63 | 0.54 | 1.73 | 0.23 | 1.62 | 0.54 |
| 840 | 1.73 | 0.22 | 1.64 | 0.55 | 1.73 | 0.22 | 1.63 | 0.55 | 1.73 | 0.23 | 1.62 | 0.55 |
| 841 | 1.73 | 0.22 | 1.64 | 0.55 | 1.73 | 0.23 | 1.63 | 0.55 | 1.73 | 0.23 | 1.62 | 0.55 |
| 842 | 1.73 | 0.22 | 1.64 | 0.55 | 1.73 | 0.23 | 1.63 | 0.55 | 1.73 | 0.23 | 1.62 | 0.55 |
| 843 | 1.73 | 0.22 | 1.64 | 0.55 | 1.73 | 0.23 | 1.63 | 0.55 | 1.73 | 0.23 | 1.62 | 0.55 |
| 844 | 1.73 | 0.22 | 1.65 | 0.56 | 1.73 | 0.23 | 1.63 | 0.56 | 1.73 | 0.23 | 1.62 | 0.56 |
| 845 | 1.73 | 0.22 | 1.65 | 0.56 | 1.73 | 0.23 | 1.63 | 0.56 | 1.73 | 0.23 | 1.62 | 0.56 |
| 846 | 1.73 | 0.22 | 1.65 | 0.56 | 1.73 | 0.23 | 1.63 | 0.56 | 1.73 | 0.23 | 1.62 | 0.56 |
| 847 | 1.73 | 0.22 | 1.65 | 0.56 | 1.73 | 0.23 | 1.63 | 0.56 | 1.73 | 0.23 | 1.62 | 0.56 |
| 848 | 1.73 | 0.22 | 1.65 | 0.57 | 1.73 | 0.23 | 1.63 | 0.56 | 1.73 | 0.23 | 1.62 | 0.56 |
| 849 | 1.73 | 0.22 | 1.65 | 0.57 | 1.73 | 0.23 | 1.63 | 0.57 | 1.73 | 0.23 | 1.62 | 0.57 |
| 850 | 1.73 | 0.22 | 1.65 | 0.57 | 1.73 | 0.23 | 1.63 | 0.57 | 1.73 | 0.23 | 1.63 | 0.57 |
| 851 | 1.73 | 0.22 | 1.65 | 0.57 | 1.73 | 0.23 | 1.63 | 0.57 | 1.73 | 0.23 | 1.63 | 0.57 |
| 852 | 1.73 | 0.22 | 1.65 | 0.57 | 1.73 | 0.23 | 1.63 | 0.57 | 1.73 | 0.23 | 1.63 | 0.57 |
| 853 | 1.73 | 0.22 | 1.65 | 0.58 | 1.73 | 0.23 | 1.63 | 0.58 | 1.73 | 0.23 | 1.63 | 0.58 |
| 854 | 1.73 | 0.23 | 1.65 | 0.58 | 1.73 | 0.23 | 1.63 | 0.58 | 1.73 | 0.23 | 1.63 | 0.58 |
| 855 | 1.73 | 0.23 | 1.65 | 0.58 | 1.73 | 0.23 | 1.63 | 0.58 | 1.73 | 0.23 | 1.63 | 0.58 |
| 856 | 1.73 | 0.23 | 1.65 | 0.58 | 1.73 | 0.23 | 1.64 | 0.58 | 1.73 | 0.23 | 1.63 | 0.58 |
| 857 | 1.73 | 0.23 | 1.65 | 0.58 | 1.73 | 0.23 | 1.64 | 0.58 | 1.73 | 0.23 | 1.63 | 0.58 |
| 858 | 1.73 | 0.23 | 1.66 | 0.59 | 1.73 | 0.23 | 1.64 | 0.59 | 1.73 | 0.23 | 1.63 | 0.59 |
| 859 | 1.73 | 0.23 | 1.66 | 0.59 | 1.73 | 0.23 | 1.64 | 0.59 | 1.73 | 0.24 | 1.63 | 0.59 |
| 860 | 1.73 | 0.23 | 1.66 | 0.59 | 1.73 | 0.23 | 1.64 | 0.59 | 1.73 | 0.24 | 1.63 | 0.59 |
| 861 | 1.73 | 0.23 | 1.66 | 0.59 | 1.73 | 0.23 | 1.64 | 0.59 | 1.73 | 0.24 | 1.63 | 0.59 |
| 862 | 1.73 | 0.23 | 1.66 | 0.59 | 1.73 | 0.23 | 1.64 | 0.60 | 1.73 | 0.24 | 1.63 | 0.60 |
| 863 | 1.73 | 0.23 | 1.66 | 0.59 | 1.73 | 0.23 | 1.64 | 0.60 | 1.73 | 0.24 | 1.63 | 0.60 |
| 864 | 1.73 | 0.23 | 1.66 | 0.60 | 1.73 | 0.24 | 1.64 | 0.60 | 1.73 | 0.24 | 1.63 | 0.60 |
| 865 | 1.73 | 0.23 | 1.66 | 0.60 | 1.73 | 0.24 | 1.64 | 0.60 | 1.73 | 0.24 | 1.63 | 0.60 |
| 866 | 1.73 | 0.23 | 1.66 | 0.60 | 1.73 | 0.24 | 1.64 | 0.60 | 1.73 | 0.24 | 1.63 | 0.60 |
| 867 | 1.73 | 0.23 | 1.66 | 0.60 | 1.73 | 0.24 | 1.64 | 0.61 | 1.73 | 0.24 | 1.63 | 0.61 |
| 868 | 1.73 | 0.23 | 1.66 | 0.60 | 1.73 | 0.24 | 1.64 | 0.61 | 1.73 | 0.24 | 1.64 | 0.61 |
| 869 | 1.73 | 0.23 | 1.66 | 0.61 | 1.73 | 0.24 | 1.64 | 0.61 | 1.73 | 0.24 | 1.64 | 0.61 |
| 870 | 1.73 | 0.23 | 1.66 | 0.61 | 1.73 | 0.24 | 1.65 | 0.61 | 1.73 | 0.24 | 1.64 | 0.61 |
| 871 | 1.73 | 0.23 | 1.67 | 0.61 | 1.73 | 0.24 | 1.65 | 0.61 | 1.73 | 0.24 | 1.64 | 0.62 |
| 872 | 1.73 | 0.23 | 1.67 | 0.61 | 1.73 | 0.24 | 1.65 | 0.62 | 1.73 | 0.24 | 1.64 | 0.62 |
| 873 | 1.73 | 0.24 | 1.67 | 0.61 | 1.73 | 0.24 | 1.65 | 0.62 | 1.73 | 0.24 | 1.64 | 0.62 |
| 874 | 1.73 | 0.24 | 1.67 | 0.62 | 1.73 | 0.24 | 1.65 | 0.62 | 1.73 | 0.24 | 1.64 | 0.62 |
| 875 | 1.73 | 0.24 | 1.67 | 0.62 | 1.73 | 0.24 | 1.65 | 0.62 | 1.73 | 0.24 | 1.64 | 0.62 |
| 876 | 1.73 | 0.24 | 1.67 | 0.62 | 1.73 | 0.24 | 1.65 | 0.63 | 1.73 | 0.24 | 1.64 | 0.63 |
| 877 | 1.73 | 0.24 | 1.67 | 0.62 | 1.73 | 0.24 | 1.65 | 0.63 | 1.73 | 0.24 | 1.64 | 0.63 |
| 878 | 1.73 | 0.24 | 1.67 | 0.62 | 1.73 | 0.24 | 1.65 | 0.63 | 1.73 | 0.24 | 1.64 | 0.63 |
| 879 | 1.73 | 0.24 | 1.67 | 0.63 | 1.73 | 0.24 | 1.65 | 0.63 | 1.73 | 0.24 | 1.64 | 0.63 |
| 880 | 1.73 | 0.24 | 1.67 | 0.63 | 1.73 | 0.24 | 1.65 | 0.63 | 1.73 | 0.24 | 1.64 | 0.64 |
| 881 | 1.73 | 0.24 | 1.67 | 0.63 | 1.73 | 0.24 | 1.65 | 0.64 | 1.73 | 0.25 | 1.65 | 0.64 |
| 882 | 1.73 | 0.24 | 1.67 | 0.63 | 1.73 | 0.24 | 1.66 | 0.64 | 1.73 | 0.25 | 1.65 | 0.64 |
| 883 | 1.73 | 0.24 | 1.68 | 0.63 | 1.73 | 0.24 | 1.66 | 0.64 | 1.73 | 0.25 | 1.65 | 0.64 |
| 884 | 1.73 | 0.24 | 1.68 | 0.64 | 1.73 | 0.25 | 1.66 | 0.64 | 1.73 | 0.25 | 1.65 | 0.65 |
| 885 | 1.73 | 0.24 | 1.68 | 0.64 | 1.73 | 0.25 | 1.66 | 0.65 | 1.73 | 0.25 | 1.65 | 0.65 |
| 886 | 1.73 | 0.24 | 1.68 | 0.64 | 1.73 | 0.25 | 1.66 | 0.65 | 1.73 | 0.25 | 1.65 | 0.65 |
| 887 | 1.73 | 0.24 | 1.68 | 0.64 | 1.73 | 0.25 | 1.66 | 0.65 | 1.73 | 0.25 | 1.65 | 0.65 |
| 888 | 1.73 | 0.24 | 1.68 | 0.65 | 1.73 | 0.25 | 1.66 | 0.65 | 1.73 | 0.25 | 1.65 | 0.65 |
| 889 | 1.73 | 0.24 | 1.68 | 0.65 | 1.73 | 0.25 | 1.66 | 0.65 | 1.73 | 0.25 | 1.65 | 0.66 |
| 890 | 1.73 | 0.24 | 1.68 | 0.65 | 1.73 | 0.25 | 1.66 | 0.66 | 1.73 | 0.25 | 1.65 | 0.66 |
| 891 | 1.73 | 0.25 | 1.68 | 0.65 | 1.73 | 0.25 | 1.66 | 0.66 | 1.73 | 0.25 | 1.65 | 0.66 |
| 892 | 1.73 | 0.25 | 1.68 | 0.65 | 1.73 | 0.25 | 1.67 | 0.66 | 1.73 | 0.25 | 1.66 | 0.66 |
| 893 | 1.73 | 0.25 | 1.69 | 0.66 | 1.73 | 0.25 | 1.67 | 0.66 | 1.73 | 0.25 | 1.66 | 0.67 |
| 894 | 1.73 | 0.25 | 1.69 | 0.66 | 1.73 | 0.25 | 1.67 | 0.67 | 1.73 | 0.25 | 1.66 | 0.67 |
| 895 | 1.73 | 0.25 | 1.69 | 0.66 | 1.73 | 0.25 | 1.67 | 0.67 | 1.73 | 0.25 | 1.66 | 0.67 |
| 896 | 1.73 | 0.25 | 1.69 | 0.66 | 1.73 | 0.25 | 1.67 | 0.67 | 1.73 | 0.25 | 1.66 | 0.67 |
| 897 | 1.73 | 0.25 | 1.69 | 0.66 | 1.73 | 0.25 | 1.67 | 0.67 | 1.73 | 0.25 | 1.66 | 0.68 |
| 898 | 1.73 | 0.25 | 1.69 | 0.67 | 1.73 | 0.25 | 1.67 | 0.68 | 1.73 | 0.25 | 1.66 | 0.68 |
| 899 | 1.73 | 0.25 | 1.69 | 0.67 | 1.73 | 0.25 | 1.67 | 0.68 | 1.73 | 0.25 | 1.66 | 0.68 |
| 900 | 1.73 | 0.25 | 1.69 | 0.67 | 1.73 | 0.25 | 1.67 | 0.68 | 1.73 | 0.26 | 1.66 | 0.68 |
| 901 | 1.73 | 0.25 | 1.69 | 0.67 | 1.73 | 0.25 | 1.68 | 0.68 | 1.73 | 0.26 | 1.67 | 0.69 |
| 902 | 1.73 | 0.25 | 1.70 | 0.68 | 1.73 | 0.26 | 1.68 | 0.68 | 1.73 | 0.26 | 1.67 | 0.69 |
| 903 | 1.73 | 0.25 | 1.70 | 0.68 | 1.73 | 0.26 | 1.68 | 0.69 | 1.73 | 0.26 | 1.67 | 0.69 |
| 904 | 1.73 | 0.25 | 1.70 | 0.68 | 1.73 | 0.26 | 1.68 | 0.69 | 1.73 | 0.26 | 1.67 | 0.69 |
| 905 | 1.74 | 0.25 | 1.70 | 0.68 | 1.73 | 0.26 | 1.68 | 0.69 | 1.73 | 0.26 | 1.67 | 0.70 |
| 906 | 1.74 | 0.25 | 1.70 | 0.69 | 1.73 | 0.26 | 1.68 | 0.69 | 1.73 | 0.26 | 1.67 | 0.70 |
| 907 | 1.74 | 0.26 | 1.70 | 0.69 | 1.73 | 0.26 | 1.68 | 0.70 | 1.73 | 0.26 | 1.67 | 0.70 |
| 908 | 1.74 | 0.26 | 1.70 | 0.69 | 1.73 | 0.26 | 1.69 | 0.70 | 1.73 | 0.26 | 1.68 | 0.70 |
| 909 | 1.74 | 0.26 | 1.71 | 0.69 | 1.73 | 0.26 | 1.69 | 0.70 | 1.73 | 0.26 | 1.68 | 0.70 |
| 910 | 1.74 | 0.26 | 1.71 | 0.69 | 1.73 | 0.26 | 1.69 | 0.70 | 1.73 | 0.26 | 1.68 | 0.71 |
| 911 | 1.74 | 0.26 | 1.71 | 0.70 | 1.73 | 0.26 | 1.69 | 0.71 | 1.73 | 0.26 | 1.68 | 0.71 |
| 912 | 1.74 | 0.26 | 1.71 | 0.70 | 1.73 | 0.26 | 1.69 | 0.71 | 1.73 | 0.26 | 1.68 | 0.71 |
| 913 | 1.74 | 0.26 | 1.71 | 0.70 | 1.73 | 0.26 | 1.69 | 0.71 | 1.73 | 0.26 | 1.68 | 0.71 |
| 914 | 1.74 | 0.26 | 1.71 | 0.70 | 1.73 | 0.26 | 1.69 | 0.71 | 1.73 | 0.26 | 1.68 | 0.72 |
| 915 | 1.74 | 0.26 | 1.71 | 0.71 | 1.73 | 0.26 | 1.70 | 0.72 | 1.73 | 0.26 | 1.69 | 0.72 |
| 916 | 1.74 | 0.26 | 1.72 | 0.71 | 1.73 | 0.26 | 1.70 | 0.72 | 1.73 | 0.27 | 1.69 | 0.72 |
| 917 | 1.74 | 0.26 | 1.72 | 0.71 | 1.73 | 0.26 | 1.70 | 0.72 | 1.73 | 0.27 | 1.69 | 0.72 |
| 918 | 1.74 | 0.26 | 1.72 | 0.71 | 1.73 | 0.27 | 1.70 | 0.72 | 1.73 | 0.27 | 1.69 | 0.73 |
| 919 | 1.74 | 0.26 | 1.72 | 0.71 | 1.74 | 0.27 | 1.70 | 0.72 | 1.73 | 0.27 | 1.69 | 0.73 |
| 920 | 1.74 | 0.26 | 1.72 | 0.72 | 1.74 | 0.27 | 1.70 | 0.73 | 1.73 | 0.27 | 1.69 | 0.73 |
| 921 | 1.74 | 0.27 | 1.73 | 0.72 | 1.74 | 0.27 | 1.71 | 0.73 | 1.73 | 0.27 | 1.70 | 0.73 |
| 922 | 1.74 | 0.27 | 1.73 | 0.72 | 1.74 | 0.27 | 1.71 | 0.73 | 1.73 | 0.27 | 1.70 | 0.74 |
| 923 | 1.74 | 0.27 | 1.73 | 0.72 | 1.74 | 0.27 | 1.71 | 0.73 | 1.73 | 0.27 | 1.70 | 0.74 |
| 924 | 1.74 | 0.27 | 1.73 | 0.73 | 1.74 | 0.27 | 1.71 | 0.74 | 1.73 | 0.27 | 1.70 | 0.74 |
| 925 | 1.74 | 0.27 | 1.73 | 0.73 | 1.74 | 0.27 | 1.71 | 0.74 | 1.73 | 0.27 | 1.70 | 0.74 |
| 926 | 1.74 | 0.27 | 1.73 | 0.73 | 1.74 | 0.27 | 1.72 | 0.74 | 1.73 | 0.27 | 1.71 | 0.75 |
| 927 | 1.74 | 0.27 | 1.74 | 0.73 | 1.74 | 0.27 | 1.72 | 0.74 | 1.73 | 0.27 | 1.71 | 0.75 |
| 928 | 1.74 | 0.27 | 1.74 | 0.73 | 1.74 | 0.27 | 1.72 | 0.75 | 1.73 | 0.27 | 1.71 | 0.75 |
| 929 | 1.74 | 0.27 | 1.74 | 0.74 | 1.74 | 0.27 | 1.72 | 0.75 | 1.73 | 0.27 | 1.71 | 0.75 |
| 930 | 1.74 | 0.27 | 1.74 | 0.74 | 1.74 | 0.27 | 1.72 | 0.75 | 1.74 | 0.27 | 1.71 | 0.75 |
| 931 | 1.74 | 0.27 | 1.75 | 0.74 | 1.74 | 0.27 | 1.73 | 0.75 | 1.74 | 0.28 | 1.72 | 0.76 |
| 932 | 1.74 | 0.27 | 1.75 | 0.74 | 1.74 | 0.28 | 1.73 | 0.75 | 1.74 | 0.28 | 1.72 | 0.76 |
| 933 | 1.74 | 0.27 | 1.75 | 0.75 | 1.74 | 0.28 | 1.73 | 0.76 | 1.74 | 0.28 | 1.72 | 0.76 |
| 934 | 1.74 | 0.27 | 1.75 | 0.75 | 1.74 | 0.28 | 1.73 | 0.76 | 1.74 | 0.28 | 1.72 | 0.76 |
| 935 | 1.74 | 0.28 | 1.75 | 0.75 | 1.74 | 0.28 | 1.74 | 0.76 | 1.74 | 0.28 | 1.72 | 0.77 |
| 936 | 1.74 | 0.28 | 1.76 | 0.75 | 1.74 | 0.28 | 1.74 | 0.76 | 1.74 | 0.28 | 1.73 | 0.77 |
| 937 | 1.74 | 0.28 | 1.76 | 0.75 | 1.74 | 0.28 | 1.74 | 0.76 | 1.74 | 0.28 | 1.73 | 0.77 |
| 938 | 1.74 | 0.28 | 1.76 | 0.76 | 1.74 | 0.28 | 1.74 | 0.77 | 1.74 | 0.28 | 1.73 | 0.77 |
| 939 | 1.74 | 0.28 | 1.76 | 0.76 | 1.74 | 0.28 | 1.75 | 0.77 | 1.74 | 0.28 | 1.73 | 0.77 |
| 940 | 1.74 | 0.28 | 1.77 | 0.76 | 1.74 | 0.28 | 1.75 | 0.77 | 1.74 | 0.28 | 1.74 | 0.78 |
| 941 | 1.74 | 0.28 | 1.77 | 0.76 | 1.74 | 0.28 | 1.75 | 0.77 | 1.74 | 0.28 | 1.74 | 0.78 |
| 942 | 1.74 | 0.28 | 1.77 | 0.76 | 1.74 | 0.28 | 1.75 | 0.78 | 1.74 | 0.28 | 1.74 | 0.78 |
| 943 | 1.74 | 0.28 | 1.77 | 0.77 | 1.74 | 0.28 | 1.75 | 0.78 | 1.74 | 0.29 | 1.74 | 0.78 |
| 944 | 1.74 | 0.28 | 1.78 | 0.77 | 1.74 | 0.29 | 1.76 | 0.78 | 1.74 | 0.29 | 1.75 | 0.78 |
| 945 | 1.74 | 0.28 | 1.78 | 0.77 | 1.74 | 0.29 | 1.76 | 0.78 | 1.74 | 0.29 | 1.75 | 0.79 |
| 946 | 1.74 | 0.29 | 1.78 | 0.77 | 1.74 | 0.29 | 1.76 | 0.78 | 1.74 | 0.29 | 1.75 | 0.79 |
| 947 | 1.74 | 0.29 | 1.78 | 0.77 | 1.74 | 0.29 | 1.77 | 0.78 | 1.74 | 0.29 | 1.75 | 0.79 |
| 948 | 1.74 | 0.29 | 1.79 | 0.77 | 1.74 | 0.29 | 1.77 | 0.79 | 1.74 | 0.29 | 1.76 | 0.79 |
| 949 | 1.74 | 0.29 | 1.79 | 0.78 | 1.74 | 0.29 | 1.77 | 0.79 | 1.74 | 0.29 | 1.76 | 0.79 |
| 950 | 1.75 | 0.29 | 1.79 | 0.78 | 1.74 | 0.29 | 1.77 | 0.79 | 1.74 | 0.29 | 1.76 | 0.80 |
| 951 | 1.75 | 0.29 | 1.79 | 0.78 | 1.74 | 0.29 | 1.78 | 0.79 | 1.74 | 0.29 | 1.76 | 0.80 |
| 952 | 1.75 | 0.29 | 1.80 | 0.78 | 1.74 | 0.29 | 1.78 | 0.79 | 1.74 | 0.29 | 1.77 | 0.80 |
| 953 | 1.75 | 0.29 | 1.80 | 0.78 | 1.74 | 0.29 | 1.78 | 0.80 | 1.74 | 0.29 | 1.77 | 0.80 |
| 954 | 1.75 | 0.29 | 1.80 | 0.78 | 1.74 | 0.29 | 1.78 | 0.80 | 1.74 | 0.30 | 1.77 | 0.80 |
| 955 | 1.75 | 0.29 | 1.81 | 0.79 | 1.74 | 0.30 | 1.79 | 0.80 | 1.74 | 0.30 | 1.77 | 0.81 |
| 956 | 1.75 | 0.29 | 1.81 | 0.79 | 1.74 | 0.30 | 1.79 | 0.80 | 1.74 | 0.30 | 1.78 | 0.81 |
| 957 | 1.75 | 0.29 | 1.81 | 0.79 | 1.74 | 0.30 | 1.79 | 0.80 | 1.74 | 0.30 | 1.78 | 0.81 |
| 958 | 1.75 | 0.30 | 1.81 | 0.79 | 1.74 | 0.30 | 1.80 | 0.80 | 1.74 | 0.30 | 1.78 | 0.81 |
| 959 | 1.75 | 0.30 | 1.82 | 0.79 | 1.74 | 0.30 | 1.80 | 0.80 | 1.74 | 0.30 | 1.79 | 0.81 |
| 960 | 1.75 | 0.30 | 1.82 | 0.79 | 1.74 | 0.30 | 1.80 | 0.81 | 1.74 | 0.30 | 1.79 | 0.81 |
| 961 | 1.75 | 0.30 | 1.82 | 0.79 | 1.75 | 0.30 | 1.80 | 0.81 | 1.74 | 0.30 | 1.79 | 0.82 |
| 962 | 1.75 | 0.30 | 1.83 | 0.80 | 1.75 | 0.30 | 1.81 | 0.81 | 1.74 | 0.30 | 1.79 | 0.82 |
| 963 | 1.75 | 0.30 | 1.83 | 0.80 | 1.75 | 0.30 | 1.81 | 0.81 | 1.74 | 0.30 | 1.80 | 0.82 |
| 964 | 1.75 | 0.30 | 1.83 | 0.80 | 1.75 | 0.30 | 1.81 | 0.81 | 1.74 | 0.30 | 1.80 | 0.82 |
| 965 | 1.75 | 0.30 | 1.84 | 0.80 | 1.75 | 0.30 | 1.82 | 0.81 | 1.74 | 0.31 | 1.80 | 0.82 |
| 966 | 1.75 | 0.30 | 1.84 | 0.80 | 1.75 | 0.31 | 1.82 | 0.81 | 1.75 | 0.31 | 1.81 | 0.82 |
| 967 | 1.75 | 0.30 | 1.84 | 0.80 | 1.75 | 0.31 | 1.82 | 0.82 | 1.75 | 0.31 | 1.81 | 0.82 |
| 968 | 1.75 | 0.31 | 1.84 | 0.80 | 1.75 | 0.31 | 1.83 | 0.82 | 1.75 | 0.31 | 1.81 | 0.83 |
| 969 | 1.75 | 0.31 | 1.85 | 0.80 | 1.75 | 0.31 | 1.83 | 0.82 | 1.75 | 0.31 | 1.82 | 0.83 |
| 970 | 1.76 | 0.31 | 1.85 | 0.80 | 1.75 | 0.31 | 1.83 | 0.82 | 1.75 | 0.31 | 1.82 | 0.83 |
| 971 | 1.76 | 0.31 | 1.85 | 0.80 | 1.75 | 0.31 | 1.83 | 0.82 | 1.75 | 0.31 | 1.82 | 0.83 |
| 972 | 1.76 | 0.31 | 1.86 | 0.81 | 1.75 | 0.31 | 1.84 | 0.82 | 1.75 | 0.31 | 1.82 | 0.83 |
| 973 | 1.76 | 0.31 | 1.86 | 0.81 | 1.75 | 0.31 | 1.84 | 0.82 | 1.75 | 0.31 | 1.83 | 0.83 |
| 974 | 1.76 | 0.31 | 1.86 | 0.81 | 1.75 | 0.31 | 1.84 | 0.82 | 1.75 | 0.31 | 1.83 | 0.83 |
| 975 | 1.76 | 0.31 | 1.87 | 0.81 | 1.75 | 0.32 | 1.85 | 0.82 | 1.75 | 0.32 | 1.83 | 0.83 |
| 976 | 1.76 | 0.31 | 1.87 | 0.81 | 1.75 | 0.32 | 1.85 | 0.82 | 1.75 | 0.32 | 1.84 | 0.83 |
| 977 | 1.76 | 0.31 | 1.87 | 0.81 | 1.75 | 0.32 | 1.85 | 0.83 | 1.75 | 0.32 | 1.84 | 0.84 |
| 978 | 1.76 | 0.32 | 1.87 | 0.81 | 1.75 | 0.32 | 1.86 | 0.83 | 1.75 | 0.32 | 1.84 | 0.84 |
| 979 | 1.76 | 0.32 | 1.88 | 0.81 | 1.76 | 0.32 | 1.86 | 0.83 | 1.75 | 0.32 | 1.85 | 0.84 |
| 980 | 1.76 | 0.32 | 1.88 | 0.81 | 1.76 | 0.32 | 1.86 | 0.83 | 1.75 | 0.32 | 1.85 | 0.84 |
| 981 | 1.76 | 0.32 | 1.88 | 0.81 | 1.76 | 0.32 | 1.86 | 0.83 | 1.75 | 0.32 | 1.85 | 0.84 |
| 982 | 1.76 | 0.32 | 1.89 | 0.81 | 1.76 | 0.32 | 1.87 | 0.83 | 1.75 | 0.32 | 1.86 | 0.84 |
| 983 | 1.77 | 0.32 | 1.89 | 0.81 | 1.76 | 0.32 | 1.87 | 0.83 | 1.76 | 0.32 | 1.86 | 0.84 |
| 984 | 1.77 | 0.32 | 1.89 | 0.81 | 1.76 | 0.32 | 1.87 | 0.83 | 1.76 | 0.33 | 1.86 | 0.84 |
| 985 | 1.77 | 0.32 | 1.90 | 0.81 | 1.76 | 0.33 | 1.88 | 0.83 | 1.76 | 0.33 | 1.86 | 0.84 |
| 986 | 1.77 | 0.32 | 1.90 | 0.81 | 1.76 | 0.33 | 1.88 | 0.83 | 1.76 | 0.33 | 1.87 | 0.84 |
| 987 | 1.77 | 0.32 | 1.90 | 0.82 | 1.76 | 0.33 | 1.88 | 0.83 | 1.76 | 0.33 | 1.87 | 0.84 |
| 988 | 1.77 | 0.33 | 1.91 | 0.82 | 1.76 | 0.33 | 1.89 | 0.83 | 1.76 | 0.33 | 1.87 | 0.84 |
| 989 | 1.77 | 0.33 | 1.91 | 0.82 | 1.76 | 0.33 | 1.89 | 0.83 | 1.76 | 0.33 | 1.88 | 0.84 |
| 990 | 1.77 | 0.33 | 1.91 | 0.82 | 1.76 | 0.33 | 1.89 | 0.83 | 1.76 | 0.33 | 1.88 | 0.85 |
| 991 | 1.77 | 0.33 | 1.91 | 0.82 | 1.77 | 0.33 | 1.90 | 0.83 | 1.76 | 0.33 | 1.88 | 0.85 |
| 992 | 1.77 | 0.33 | 1.92 | 0.82 | 1.77 | 0.33 | 1.90 | 0.83 | 1.76 | 0.33 | 1.89 | 0.85 |
| 993 | 1.77 | 0.33 | 1.92 | 0.82 | 1.77 | 0.33 | 1.90 | 0.84 | 1.76 | 0.34 | 1.89 | 0.85 |
| 994 | 1.78 | 0.33 | 1.92 | 0.82 | 1.77 | 0.34 | 1.91 | 0.84 | 1.77 | 0.34 | 1.89 | 0.85 |
| 995 | 1.78 | 0.33 | 1.93 | 0.82 | 1.77 | 0.34 | 1.91 | 0.84 | 1.77 | 0.34 | 1.90 | 0.85 |
| 996 | 1.78 | 0.33 | 1.93 | 0.82 | 1.77 | 0.34 | 1.91 | 0.84 | 1.77 | 0.34 | 1.90 | 0.85 |
| 997 | 1.78 | 0.33 | 1.93 | 0.82 | 1.77 | 0.34 | 1.91 | 0.84 | 1.77 | 0.34 | 1.90 | 0.85 |
| 998 | 1.78 | 0.33 | 1.94 | 0.82 | 1.77 | 0.34 | 1.92 | 0.84 | 1.77 | 0.34 | 1.91 | 0.85 |
| 999 | 1.78 | 0.34 | 1.94 | 0.82 | 1.77 | 0.34 | 1.92 | 0.84 | 1.77 | 0.34 | 1.91 | 0.85 |
| 1000 | 1.78 | 0.34 | 1.94 | 0.82 | 1.78 | 0.34 | 1.92 | 0.84 | 1.77 | 0.34 | 1.91 | 0.85 |
| 1006 | 1.79 | 0.34 | 1.96 | 0.82 | 1.78 | 0.35 | 1.94 | 0.84 | 1.78 | 0.35 | 1.93 | 0.85 |
| 1012 | 1.80 | 0.35 | 1.98 | 0.82 | 1.79 | 0.35 | 1.96 | 0.84 | 1.79 | 0.35 | 1.95 | 0.85 |
| 1018 | 1.81 | 0.35 | 1.99 | 0.81 | 1.80 | 0.36 | 1.98 | 0.83 | 1.80 | 0.36 | 1.97 | 0.85 |
| 1024 | 1.82 | 0.35 | 2.01 | 0.81 | 1.81 | 0.36 | 2.00 | 0.83 | 1.81 | 0.36 | 1.99 | 0.85 |
| 1030 | 1.83 | 0.35 | 2.03 | 0.81 | 1.82 | 0.36 | 2.01 | 0.83 | 1.82 | 0.37 | 2.00 | 0.84 |
| 1036 | 1.84 | 0.36 | 2.04 | 0.80 | 1.83 | 0.36 | 2.03 | 0.82 | 1.83 | 0.37 | 2.02 | 0.84 |
| 1042 | 1.85 | 0.36 | 2.06 | 0.79 | 1.84 | 0.37 | 2.05 | 0.82 | 1.84 | 0.37 | 2.04 | 0.84 |
| 1048 | 1.86 | 0.35 | 2.07 | 0.79 | 1.85 | 0.36 | 2.06 | 0.81 | 1.85 | 0.37 | 2.05 | 0.83 |
| 1054 | 1.87 | 0.35 | 2.09 | 0.78 | 1.86 | 0.36 | 2.08 | 0.81 | 1.86 | 0.37 | 2.07 | 0.82 |
| 1060 | 1.88 | 0.35 | 2.10 | 0.77 | 1.87 | 0.36 | 2.09 | 0.80 | 1.87 | 0.36 | 2.09 | 0.81 |
| 1066 | 1.88 | 0.35 | 2.12 | 0.76 | 1.88 | 0.36 | 2.11 | 0.79 | 1.88 | 0.36 | 2.10 | 0.81 |
| 1072 | 1.89 | 0.34 | 2.13 | 0.75 | 1.89 | 0.35 | 2.12 | 0.78 | 1.89 | 0.36 | 2.12 | 0.80 |
| 1078 | 1.90 | 0.34 | 2.14 | 0.74 | 1.90 | 0.35 | 2.13 | 0.77 | 1.90 | 0.35 | 2.13 | 0.79 |
| 1084 | 1.91 | 0.33 | 2.15 | 0.73 | 1.91 | 0.34 | 2.14 | 0.76 | 1.91 | 0.34 | 2.14 | 0.77 |
| 1090 | 1.91 | 0.33 | 2.16 | 0.72 | 1.91 | 0.33 | 2.16 | 0.75 | 1.91 | 0.34 | 2.15 | 0.76 |
| 1096 | 1.92 | 0.32 | 2.17 | 0.71 | 1.92 | 0.33 | 2.17 | 0.74 | 1.92 | 0.33 | 2.16 | 0.75 |
| 1102 | 1.92 | 0.31 | 2.18 | 0.70 | 1.92 | 0.32 | 2.17 | 0.72 | 1.92 | 0.32 | 2.17 | 0.74 |
| 1108 | 1.93 | 0.31 | 2.19 | 0.69 | 1.93 | 0.31 | 2.18 | 0.71 | 1.93 | 0.32 | 2.18 | 0.72 |
| 1114 | 1.93 | 0.30 | 2.19 | 0.67 | 1.93 | 0.31 | 2.19 | 0.70 | 1.93 | 0.31 | 2.18 | 0.71 |
| 1120 | 1.93 | 0.30 | 2.20 | 0.66 | 1.93 | 0.30 | 2.19 | 0.69 | 1.93 | 0.30 | 2.19 | 0.70 |
| 1126 | 1.93 | 0.29 | 2.20 | 0.65 | 1.94 | 0.30 | 2.20 | 0.67 | 1.94 | 0.30 | 2.19 | 0.68 |
| 1132 | 1.94 | 0.29 | 2.21 | 0.64 | 1.94 | 0.29 | 2.20 | 0.66 | 1.94 | 0.29 | 2.20 | 0.67 |
| 1138 | 1.94 | 0.28 | 2.21 | 0.62 | 1.94 | 0.28 | 2.20 | 0.65 | 1.94 | 0.29 | 2.20 | 0.66 |
| 1144 | 1.94 | 0.28 | 2.21 | 0.61 | 1.94 | 0.28 | 2.20 | 0.64 | 1.94 | 0.28 | 2.20 | 0.65 |
| 1150 | 1.94 | 0.27 | 2.21 | 0.60 | 1.94 | 0.27 | 2.20 | 0.62 | 1.94 | 0.28 | 2.20 | 0.64 |
| 1156 | 1.94 | 0.27 | 2.21 | 0.59 | 1.94 | 0.27 | 2.20 | 0.61 | 1.94 | 0.27 | 2.20 | 0.62 |
| 1162 | 1.94 | 0.26 | 2.21 | 0.58 | 1.94 | 0.27 | 2.20 | 0.60 | 1.94 | 0.27 | 2.20 | 0.61 |
| 1168 | 1.94 | 0.26 | 2.21 | 0.57 | 1.94 | 0.26 | 2.20 | 0.59 | 1.94 | 0.26 | 2.19 | 0.60 |
| 1174 | 1.94 | 0.26 | 2.21 | 0.56 | 1.94 | 0.26 | 2.20 | 0.58 | 1.94 | 0.26 | 2.19 | 0.60 |
| 1180 | 1.94 | 0.25 | 2.20 | 0.55 | 1.94 | 0.25 | 2.20 | 0.57 | 1.94 | 0.26 | 2.19 | 0.59 |
| 1186 | 1.94 | 0.25 | 2.20 | 0.55 | 1.94 | 0.25 | 2.19 | 0.57 | 1.94 | 0.25 | 2.18 | 0.58 |
| 1192 | 1.94 | 0.24 | 2.20 | 0.54 | 1.94 | 0.25 | 2.19 | 0.56 | 1.94 | 0.25 | 2.18 | 0.57 |
| 1198 | 1.94 | 0.24 | 2.19 | 0.53 | 1.94 | 0.24 | 2.19 | 0.55 | 1.94 | 0.25 | 2.18 | 0.57 |
| 1204 | 1.94 | 0.24 | 2.19 | 0.52 | 1.94 | 0.24 | 2.18 | 0.55 | 1.94 | 0.24 | 2.17 | 0.56 |
| 1210 | 1.94 | 0.24 | 2.19 | 0.52 | 1.94 | 0.24 | 2.18 | 0.54 | 1.94 | 0.24 | 2.17 | 0.55 |
| 1216 | 1.94 | 0.23 | 2.18 | 0.51 | 1.94 | 0.24 | 2.17 | 0.54 | 1.94 | 0.24 | 2.16 | 0.55 |
| 1222 | 1.94 | 0.23 | 2.18 | 0.51 | 1.94 | 0.23 | 2.17 | 0.53 | 1.94 | 0.23 | 2.16 | 0.55 |
| 1228 | 1.94 | 0.23 | 2.17 | 0.50 | 1.94 | 0.23 | 2.16 | 0.53 | 1.94 | 0.23 | 2.15 | 0.54 |
| 1234 | 1.94 | 0.23 | 2.17 | 0.50 | 1.94 | 0.23 | 2.16 | 0.52 | 1.94 | 0.23 | 2.15 | 0.54 |
| 1240 | 1.94 | 0.22 | 2.16 | 0.50 | 1.94 | 0.23 | 2.15 | 0.52 | 1.93 | 0.23 | 2.14 | 0.54 |
| 1246 | 1.94 | 0.22 | 2.16 | 0.49 | 1.94 | 0.22 | 2.15 | 0.52 | 1.93 | 0.23 | 2.14 | 0.54 |
| 1252 | 1.94 | 0.22 | 2.16 | 0.49 | 1.93 | 0.22 | 2.14 | 0.52 | 1.93 | 0.22 | 2.13 | 0.53 |
| 1258 | 1.94 | 0.22 | 2.15 | 0.49 | 1.93 | 0.22 | 2.14 | 0.51 | 1.93 | 0.22 | 2.13 | 0.53 |
| 1264 | 1.93 | 0.22 | 2.15 | 0.49 | 1.93 | 0.22 | 2.14 | 0.51 | 1.93 | 0.22 | 2.12 | 0.53 |
| 1270 | 1.93 | 0.22 | 2.14 | 0.49 | 1.93 | 0.22 | 2.13 | 0.51 | 1.93 | 0.22 | 2.12 | 0.53 |
| 1276 | 1.93 | 0.21 | 2.14 | 0.48 | 1.93 | 0.22 | 2.13 | 0.51 | 1.93 | 0.22 | 2.11 | 0.53 |
| 1282 | 1.93 | 0.21 | 2.14 | 0.48 | 1.93 | 0.21 | 2.12 | 0.51 | 1.93 | 0.22 | 2.11 | 0.53 |
| 1288 | 1.93 | 0.21 | 2.13 | 0.48 | 1.93 | 0.21 | 2.12 | 0.51 | 1.93 | 0.22 | 2.10 | 0.53 |
| 1294 | 1.93 | 0.21 | 2.13 | 0.48 | 1.93 | 0.21 | 2.11 | 0.51 | 1.93 | 0.21 | 2.10 | 0.54 |
| 1300 | 1.93 | 0.21 | 2.12 | 0.48 | 1.93 | 0.21 | 2.11 | 0.51 | 1.93 | 0.21 | 2.10 | 0.54 |
| 1306 | 1.93 | 0.21 | 2.12 | 0.48 | 1.93 | 0.21 | 2.11 | 0.52 | 1.93 | 0.21 | 2.09 | 0.54 |
| 1312 | 1.93 | 0.21 | 2.12 | 0.48 | 1.93 | 0.21 | 2.10 | 0.52 | 1.92 | 0.21 | 2.09 | 0.54 |
| 1318 | 1.93 | 0.21 | 2.11 | 0.49 | 1.93 | 0.21 | 2.10 | 0.52 | 1.92 | 0.21 | 2.09 | 0.54 |
| 1324 | 1.93 | 0.20 | 2.11 | 0.49 | 1.92 | 0.21 | 2.10 | 0.52 | 1.92 | 0.21 | 2.08 | 0.55 |
| 1330 | 1.93 | 0.20 | 2.11 | 0.49 | 1.92 | 0.21 | 2.09 | 0.52 | 1.92 | 0.21 | 2.08 | 0.55 |
| 1336 | 1.93 | 0.20 | 2.11 | 0.49 | 1.92 | 0.21 | 2.09 | 0.53 | 1.92 | 0.21 | 2.08 | 0.55 |
| 1342 | 1.92 | 0.20 | 2.10 | 0.49 | 1.92 | 0.20 | 2.09 | 0.53 | 1.92 | 0.21 | 2.07 | 0.56 |
| 1348 | 1.92 | 0.20 | 2.10 | 0.49 | 1.92 | 0.20 | 2.09 | 0.53 | 1.92 | 0.21 | 2.07 | 0.56 |
| 1354 | 1.92 | 0.20 | 2.10 | 0.49 | 1.92 | 0.20 | 2.08 | 0.53 | 1.92 | 0.21 | 2.07 | 0.57 |
| 1360 | 1.92 | 0.20 | 2.10 | 0.50 | 1.92 | 0.20 | 2.08 | 0.54 | 1.92 | 0.21 | 2.07 | 0.57 |
| 1366 | 1.92 | 0.20 | 2.10 | 0.50 | 1.92 | 0.20 | 2.08 | 0.54 | 1.92 | 0.21 | 2.07 | 0.57 |
| 1372 | 1.92 | 0.20 | 2.09 | 0.50 | 1.92 | 0.20 | 2.08 | 0.55 | 1.92 | 0.21 | 2.07 | 0.58 |
| 1378 | 1.92 | 0.20 | 2.09 | 0.50 | 1.92 | 0.20 | 2.08 | 0.55 | 1.92 | 0.21 | 2.06 | 0.58 |
| 1384 | 1.92 | 0.20 | 2.09 | 0.51 | 1.92 | 0.20 | 2.08 | 0.55 | 1.92 | 0.21 | 2.06 | 0.59 |
| 1390 | 1.92 | 0.20 | 2.09 | 0.51 | 1.92 | 0.20 | 2.08 | 0.56 | 1.91 | 0.21 | 2.06 | 0.59 |
| 1396 | 1.92 | 0.20 | 2.09 | 0.51 | 1.91 | 0.20 | 2.08 | 0.56 | 1.91 | 0.21 | 2.06 | 0.60 |
| 1402 | 1.92 | 0.20 | 2.09 | 0.51 | 1.91 | 0.20 | 2.08 | 0.56 | 1.91 | 0.21 | 2.06 | 0.60 |
| 1408 | 1.92 | 0.20 | 2.09 | 0.52 | 1.91 | 0.20 | 2.08 | 0.57 | 1.91 | 0.21 | 2.06 | 0.61 |
| 1414 | 1.92 | 0.20 | 2.09 | 0.52 | 1.91 | 0.20 | 2.08 | 0.57 | 1.91 | 0.21 | 2.06 | 0.61 |
| 1420 | 1.92 | 0.20 | 2.09 | 0.52 | 1.91 | 0.20 | 2.08 | 0.58 | 1.91 | 0.21 | 2.07 | 0.62 |
| 1426 | 1.91 | 0.20 | 2.09 | 0.53 | 1.91 | 0.20 | 2.08 | 0.58 | 1.91 | 0.21 | 2.07 | 0.62 |
| 1432 | 1.91 | 0.20 | 2.09 | 0.53 | 1.91 | 0.20 | 2.08 | 0.59 | 1.91 | 0.21 | 2.07 | 0.63 |
| 1438 | 1.91 | 0.20 | 2.09 | 0.53 | 1.91 | 0.20 | 2.08 | 0.59 | 1.91 | 0.21 | 2.07 | 0.64 |
| 1444 | 1.91 | 0.20 | 2.09 | 0.53 | 1.91 | 0.20 | 2.08 | 0.59 | 1.91 | 0.21 | 2.07 | 0.64 |
| 1450 | 1.91 | 0.20 | 2.09 | 0.54 | 1.91 | 0.20 | 2.08 | 0.60 | 1.91 | 0.21 | 2.07 | 0.65 |
| 1456 | 1.91 | 0.20 | 2.09 | 0.54 | 1.91 | 0.20 | 2.09 | 0.60 | 1.91 | 0.21 | 2.08 | 0.65 |
| 1462 | 1.91 | 0.20 | 2.09 | 0.54 | 1.91 | 0.20 | 2.09 | 0.61 | 1.91 | 0.21 | 2.08 | 0.66 |
| 1468 | 1.91 | 0.20 | 2.10 | 0.55 | 1.91 | 0.20 | 2.09 | 0.61 | 1.91 | 0.21 | 2.08 | 0.66 |
| 1474 | 1.91 | 0.20 | 2.10 | 0.55 | 1.91 | 0.21 | 2.09 | 0.62 | 1.91 | 0.21 | 2.09 | 0.67 |
| 1480 | 1.91 | 0.20 | 2.10 | 0.55 | 1.91 | 0.21 | 2.10 | 0.62 | 1.91 | 0.21 | 2.09 | 0.67 |
| 1486 | 1.91 | 0.20 | 2.10 | 0.56 | 1.91 | 0.21 | 2.10 | 0.62 | 1.91 | 0.21 | 2.09 | 0.68 |
| 1492 | 1.91 | 0.20 | 2.10 | 0.56 | 1.91 | 0.21 | 2.10 | 0.63 | 1.91 | 0.21 | 2.10 | 0.68 |
| 1498 | 1.91 | 0.20 | 2.11 | 0.56 | 1.91 | 0.21 | 2.11 | 0.63 | 1.91 | 0.21 | 2.10 | 0.69 |
| 1504 | 1.91 | 0.20 | 2.11 | 0.56 | 1.91 | 0.21 | 2.11 | 0.64 | 1.91 | 0.21 | 2.11 | 0.69 |
| 1510 | 1.91 | 0.20 | 2.11 | 0.57 | 1.91 | 0.21 | 2.11 | 0.64 | 1.91 | 0.21 | 2.11 | 0.70 |
| 1516 | 1.91 | 0.20 | 2.11 | 0.57 | 1.91 | 0.21 | 2.12 | 0.64 | 1.91 | 0.21 | 2.12 | 0.70 |
| 1522 | 1.91 | 0.21 | 2.11 | 0.57 | 1.91 | 0.21 | 2.12 | 0.65 | 1.91 | 0.21 | 2.12 | 0.71 |
| 1528 | 1.91 | 0.21 | 2.12 | 0.57 | 1.91 | 0.21 | 2.12 | 0.65 | 1.91 | 0.21 | 2.13 | 0.71 |
| 1534 | 1.91 | 0.21 | 2.12 | 0.57 | 1.91 | 0.21 | 2.13 | 0.65 | 1.91 | 0.21 | 2.13 | 0.71 |
| 1540 | 1.91 | 0.21 | 2.12 | 0.58 | 1.91 | 0.21 | 2.13 | 0.66 | 1.91 | 0.22 | 2.14 | 0.72 |
| 1546 | 1.91 | 0.21 | 2.13 | 0.58 | 1.91 | 0.21 | 2.14 | 0.66 | 1.91 | 0.22 | 2.14 | 0.72 |
| 1552 | 1.91 | 0.21 | 2.13 | 0.58 | 1.91 | 0.22 | 2.14 | 0.66 | 1.91 | 0.22 | 2.15 | 0.72 |
| 1558 | 1.91 | 0.21 | 2.13 | 0.58 | 1.91 | 0.22 | 2.15 | 0.66 | 1.91 | 0.22 | 2.15 | 0.73 |
| 1564 | 1.91 | 0.21 | 2.14 | 0.58 | 1.91 | 0.22 | 2.15 | 0.67 | 1.91 | 0.22 | 2.16 | 0.73 |
| 1570 | 1.91 | 0.21 | 2.14 | 0.59 | 1.91 | 0.22 | 2.16 | 0.67 | 1.91 | 0.22 | 2.17 | 0.73 |
| 1576 | 1.91 | 0.21 | 2.14 | 0.59 | 1.91 | 0.22 | 2.16 | 0.67 | 1.91 | 0.22 | 2.17 | 0.74 |
| 1582 | 1.91 | 0.21 | 2.15 | 0.59 | 1.91 | 0.22 | 2.17 | 0.67 | 1.91 | 0.22 | 2.18 | 0.74 |
| 1588 | 1.91 | 0.22 | 2.15 | 0.59 | 1.91 | 0.22 | 2.17 | 0.68 | 1.91 | 0.22 | 2.19 | 0.74 |
| 1594 | 1.91 | 0.22 | 2.15 | 0.59 | 1.91 | 0.22 | 2.18 | 0.68 | 1.91 | 0.22 | 2.19 | 0.74 |
| 1600 | 1.91 | 0.22 | 2.16 | 0.59 | 1.91 | 0.22 | 2.19 | 0.68 | 1.91 | 0.22 | 2.20 | 0.75 |
| 1606 | 1.91 | 0.22 | 2.16 | 0.59 | 1.91 | 0.22 | 2.19 | 0.68 | 1.91 | 0.23 | 2.21 | 0.75 |
| 1612 | 1.91 | 0.22 | 2.16 | 0.59 | 1.91 | 0.23 | 2.20 | 0.68 | 1.91 | 0.23 | 2.21 | 0.75 |
| 1618 | 1.91 | 0.22 | 2.17 | 0.59 | 1.91 | 0.23 | 2.20 | 0.68 | 1.91 | 0.23 | 2.22 | 0.75 |
| 1624 | 1.91 | 0.22 | 2.17 | 0.59 | 1.91 | 0.23 | 2.21 | 0.68 | 1.91 | 0.23 | 2.23 | 0.75 |
| 1630 | 1.91 | 0.22 | 2.18 | 0.59 | 1.91 | 0.23 | 2.21 | 0.68 | 1.91 | 0.23 | 2.24 | 0.75 |
| 1636 | 1.91 | 0.22 | 2.18 | 0.59 | 1.91 | 0.23 | 2.22 | 0.68 | 1.91 | 0.23 | 2.24 | 0.75 |
| 1642 | 1.91 | 0.23 | 2.18 | 0.59 | 1.91 | 0.23 | 2.23 | 0.68 | 1.92 | 0.23 | 2.25 | 0.75 |
| 1648 | 1.91 | 0.23 | 2.19 | 0.59 | 1.91 | 0.23 | 2.23 | 0.68 | 1.92 | 0.23 | 2.26 | 0.75 |
| 1654 | 1.91 | 0.23 | 2.19 | 0.59 | 1.91 | 0.23 | 2.24 | 0.68 | 1.92 | 0.23 | 2.26 | 0.75 |
| 1660 | 1.91 | 0.23 | 2.20 | 0.59 | 1.92 | 0.23 | 2.24 | 0.68 | 1.92 | 0.23 | 2.27 | 0.75 |
| 1666 | 1.91 | 0.23 | 2.20 | 0.59 | 1.92 | 0.23 | 2.25 | 0.68 | 1.92 | 0.23 | 2.28 | 0.75 |
| 1672 | 1.91 | 0.23 | 2.20 | 0.59 | 1.92 | 0.24 | 2.25 | 0.68 | 1.92 | 0.23 | 2.29 | 0.75 |
| 1678 | 1.91 | 0.23 | 2.21 | 0.59 | 1.92 | 0.24 | 2.26 | 0.68 | 1.92 | 0.24 | 2.29 | 0.75 |
| 1684 | 1.91 | 0.23 | 2.21 | 0.59 | 1.92 | 0.24 | 2.27 | 0.68 | 1.92 | 0.24 | 2.30 | 0.75 |
| 1690 | 1.91 | 0.23 | 2.21 | 0.59 | 1.92 | 0.24 | 2.27 | 0.68 | 1.92 | 0.24 | 2.31 | 0.75 |
| 1696 | 1.92 | 0.24 | 2.22 | 0.59 | 1.92 | 0.24 | 2.28 | 0.68 | 1.92 | 0.24 | 2.31 | 0.75 |
| 1702 | 1.92 | 0.24 | 2.22 | 0.59 | 1.92 | 0.24 | 2.28 | 0.68 | 1.93 | 0.24 | 2.32 | 0.75 |
| 1708 | 1.92 | 0.24 | 2.22 | 0.59 | 1.92 | 0.24 | 2.29 | 0.68 | 1.93 | 0.24 | 2.33 | 0.74 |
| 1714 | 1.92 | 0.24 | 2.23 | 0.59 | 1.93 | 0.24 | 2.29 | 0.67 | 1.93 | 0.24 | 2.33 | 0.74 |
| 1720 | 1.92 | 0.24 | 2.23 | 0.58 | 1.93 | 0.24 | 2.30 | 0.67 | 1.93 | 0.24 | 2.34 | 0.74 |
| 1726 | 1.92 | 0.24 | 2.24 | 0.58 | 1.93 | 0.24 | 2.30 | 0.67 | 1.93 | 0.24 | 2.35 | 0.74 |
| 1732 | 1.92 | 0.24 | 2.24 | 0.58 | 1.93 | 0.25 | 2.31 | 0.67 | 1.93 | 0.24 | 2.35 | 0.74 |
| 1738 | 1.92 | 0.24 | 2.24 | 0.58 | 1.93 | 0.25 | 2.31 | 0.67 | 1.93 | 0.24 | 2.36 | 0.73 |
| 1744 | 1.92 | 0.24 | 2.25 | 0.58 | 1.93 | 0.25 | 2.32 | 0.66 | 1.93 | 0.24 | 2.36 | 0.73 |
| 1750 | 1.93 | 0.25 | 2.25 | 0.58 | 1.93 | 0.25 | 2.32 | 0.66 | 1.94 | 0.24 | 2.37 | 0.73 |
| 1756 | 1.93 | 0.25 | 2.25 | 0.57 | 1.94 | 0.25 | 2.33 | 0.66 | 1.94 | 0.24 | 2.38 | 0.72 |
| 1762 | 1.93 | 0.25 | 2.25 | 0.57 | 1.94 | 0.25 | 2.33 | 0.66 | 1.94 | 0.25 | 2.38 | 0.72 |
| 1768 | 1.93 | 0.25 | 2.26 | 0.57 | 1.94 | 0.25 | 2.34 | 0.65 | 1.94 | 0.25 | 2.39 | 0.72 |
| 1774 | 1.93 | 0.25 | 2.26 | 0.57 | 1.94 | 0.25 | 2.34 | 0.65 | 1.94 | 0.25 | 2.39 | 0.72 |
| 1780 | 1.93 | 0.25 | 2.26 | 0.57 | 1.94 | 0.25 | 2.35 | 0.65 | 1.94 | 0.25 | 2.40 | 0.71 |
| 1786 | 1.93 | 0.25 | 2.27 | 0.56 | 1.94 | 0.25 | 2.35 | 0.64 | 1.94 | 0.25 | 2.40 | 0.71 |
| 1792 | 1.94 | 0.25 | 2.27 | 0.56 | 1.94 | 0.25 | 2.35 | 0.64 | 1.95 | 0.25 | 2.41 | 0.71 |
| 1798 | 1.94 | 0.25 | 2.27 | 0.56 | 1.95 | 0.25 | 2.36 | 0.64 | 1.95 | 0.25 | 2.41 | 0.70 |
| 1804 | 1.94 | 0.25 | 2.27 | 0.56 | 1.95 | 0.25 | 2.36 | 0.63 | 1.95 | 0.25 | 2.42 | 0.70 |
| 1810 | 1.94 | 0.25 | 2.28 | 0.56 | 1.95 | 0.25 | 2.37 | 0.63 | 1.95 | 0.25 | 2.42 | 0.69 |
| 1816 | 1.94 | 0.26 | 2.28 | 0.55 | 1.95 | 0.25 | 2.37 | 0.63 | 1.95 | 0.25 | 2.43 | 0.69 |
| 1822 | 1.94 | 0.26 | 2.28 | 0.55 | 1.95 | 0.25 | 2.37 | 0.62 | 1.95 | 0.25 | 2.43 | 0.69 |
| 1828 | 1.95 | 0.26 | 2.28 | 0.55 | 1.96 | 0.25 | 2.38 | 0.62 | 1.96 | 0.25 | 2.44 | 0.68 |
| 1834 | 1.95 | 0.26 | 2.29 | 0.55 | 1.96 | 0.25 | 2.38 | 0.62 | 1.96 | 0.25 | 2.44 | 0.68 |
| 1840 | 1.95 | 0.26 | 2.29 | 0.54 | 1.96 | 0.25 | 2.38 | 0.61 | 1.96 | 0.25 | 2.45 | 0.67 |
| 1846 | 1.95 | 0.26 | 2.29 | 0.54 | 1.96 | 0.25 | 2.39 | 0.61 | 1.96 | 0.25 | 2.45 | 0.67 |
| 1852 | 1.95 | 0.26 | 2.29 | 0.54 | 1.96 | 0.25 | 2.39 | 0.61 | 1.96 | 0.25 | 2.45 | 0.67 |
| 1858 | 1.95 | 0.26 | 2.30 | 0.53 | 1.96 | 0.25 | 2.39 | 0.60 | 1.96 | 0.25 | 2.46 | 0.66 |
| 1864 | 1.96 | 0.26 | 2.30 | 0.53 | 1.97 | 0.25 | 2.40 | 0.60 | 1.97 | 0.25 | 2.46 | 0.66 |
| 1870 | 1.96 | 0.26 | 2.30 | 0.53 | 1.97 | 0.25 | 2.40 | 0.60 | 1.97 | 0.25 | 2.47 | 0.65 |
| 1876 | 1.96 | 0.26 | 2.30 | 0.53 | 1.97 | 0.25 | 2.40 | 0.59 | 1.97 | 0.25 | 2.47 | 0.65 |
| 1882 | 1.96 | 0.26 | 2.30 | 0.52 | 1.97 | 0.25 | 2.40 | 0.59 | 1.97 | 0.25 | 2.47 | 0.64 |
| 1888 | 1.96 | 0.26 | 2.30 | 0.52 | 1.97 | 0.25 | 2.41 | 0.59 | 1.97 | 0.25 | 2.48 | 0.64 |
| 1894 | 1.97 | 0.26 | 2.31 | 0.52 | 1.98 | 0.25 | 2.41 | 0.58 | 1.97 | 0.25 | 2.48 | 0.64 |
| 1900 | 1.97 | 0.26 | 2.31 | 0.52 | 1.98 | 0.25 | 2.41 | 0.58 | 1.98 | 0.25 | 2.48 | 0.63 |
| 1906 | 1.97 | 0.26 | 2.31 | 0.51 | 1.98 | 0.25 | 2.41 | 0.57 | 1.98 | 0.25 | 2.49 | 0.63 |
| 1912 | 1.97 | 0.26 | 2.31 | 0.51 | 1.98 | 0.25 | 2.42 | 0.57 | 1.98 | 0.25 | 2.49 | 0.62 |
| 1918 | 1.97 | 0.26 | 2.31 | 0.51 | 1.98 | 0.25 | 2.42 | 0.57 | 1.98 | 0.25 | 2.49 | 0.62 |
| 1924 | 1.98 | 0.26 | 2.31 | 0.50 | 1.98 | 0.25 | 2.42 | 0.56 | 1.98 | 0.25 | 2.49 | 0.61 |
| 1930 | 1.98 | 0.26 | 2.31 | 0.50 | 1.99 | 0.25 | 2.42 | 0.56 | 1.98 | 0.25 | 2.50 | 0.61 |
| 1936 | 1.98 | 0.26 | 2.31 | 0.50 | 1.99 | 0.25 | 2.42 | 0.55 | 1.99 | 0.25 | 2.50 | 0.60 |
| 1942 | 1.98 | 0.26 | 2.32 | 0.50 | 1.99 | 0.25 | 2.43 | 0.55 | 1.99 | 0.25 | 2.50 | 0.60 |
| 1948 | 1.98 | 0.26 | 2.32 | 0.49 | 1.99 | 0.25 | 2.43 | 0.55 | 1.99 | 0.24 | 2.50 | 0.60 |
| 1954 | 1.98 | 0.26 | 2.32 | 0.49 | 1.99 | 0.25 | 2.43 | 0.54 | 1.99 | 0.24 | 2.50 | 0.59 |
| 1960 | 1.99 | 0.26 | 2.32 | 0.49 | 1.99 | 0.25 | 2.43 | 0.54 | 1.99 | 0.24 | 2.51 | 0.59 |
| 1966 | 1.99 | 0.26 | 2.32 | 0.48 | 2.00 | 0.25 | 2.43 | 0.54 | 1.99 | 0.24 | 2.51 | 0.58 |
| 1972 | 1.99 | 0.26 | 2.32 | 0.48 | 2.00 | 0.25 | 2.43 | 0.53 | 1.99 | 0.24 | 2.51 | 0.58 |
| 1978 | 1.99 | 0.26 | 2.32 | 0.48 | 2.00 | 0.25 | 2.43 | 0.53 | 2.00 | 0.24 | 2.51 | 0.57 |
| 1984 | 1.99 | 0.26 | 2.32 | 0.48 | 2.00 | 0.25 | 2.44 | 0.52 | 2.00 | 0.24 | 2.51 | 0.57 |
| 1990 | 2.00 | 0.26 | 2.32 | 0.47 | 2.00 | 0.25 | 2.44 | 0.52 | 2.00 | 0.24 | 2.52 | 0.56 |
| 1996 | 2.00 | 0.26 | 2.32 | 0.47 | 2.00 | 0.25 | 2.44 | 0.52 | 2.00 | 0.24 | 2.52 | 0.56 |
| 2002 | 2.00 | 0.26 | 2.32 | 0.47 | 2.01 | 0.24 | 2.44 | 0.51 | 2.00 | 0.24 | 2.52 | 0.56 |
| 2008 | 2.00 | 0.26 | 2.32 | 0.46 | 2.01 | 0.24 | 2.44 | 0.51 | 2.00 | 0.24 | 2.52 | 0.55 |
| 2014 | 2.00 | 0.26 | 2.32 | 0.46 | 2.01 | 0.24 | 2.44 | 0.51 | 2.01 | 0.24 | 2.52 | 0.55 |
| 2020 | 2.01 | 0.26 | 2.32 | 0.46 | 2.01 | 0.24 | 2.44 | 0.50 | 2.01 | 0.24 | 2.52 | 0.54 |
| 2026 | 2.01 | 0.26 | 2.32 | 0.46 | 2.01 | 0.24 | 2.44 | 0.50 | 2.01 | 0.24 | 2.52 | 0.54 |
| 2032 | 2.01 | 0.25 | 2.32 | 0.45 | 2.01 | 0.24 | 2.44 | 0.50 | 2.01 | 0.24 | 2.52 | 0.54 |
| 2038 | 2.01 | 0.25 | 2.32 | 0.45 | 2.02 | 0.24 | 2.44 | 0.49 | 2.01 | 0.23 | 2.52 | 0.53 |
| 2044 | 2.01 | 0.25 | 2.32 | 0.45 | 2.02 | 0.24 | 2.44 | 0.49 | 2.01 | 0.23 | 2.53 | 0.53 |
| 2050 | 2.01 | 0.25 | 2.32 | 0.45 | 2.02 | 0.24 | 2.44 | 0.49 | 2.01 | 0.23 | 2.53 | 0.52 |
| 2056 | 2.02 | 0.25 | 2.32 | 0.44 | 2.02 | 0.24 | 2.45 | 0.48 | 2.01 | 0.23 | 2.53 | 0.52 |
| 2062 | 2.02 | 0.25 | 2.32 | 0.44 | 2.02 | 0.24 | 2.45 | 0.48 | 2.02 | 0.23 | 2.53 | 0.52 |
| 2068 | 2.02 | 0.25 | 2.32 | 0.44 | 2.02 | 0.24 | 2.45 | 0.47 | 2.02 | 0.23 | 2.53 | 0.51 |
| 2074 | 2.02 | 0.25 | 2.32 | 0.43 | 2.02 | 0.23 | 2.45 | 0.47 | 2.02 | 0.23 | 2.53 | 0.51 |
| 2080 | 2.02 | 0.25 | 2.32 | 0.43 | 2.03 | 0.23 | 2.45 | 0.47 | 2.02 | 0.23 | 2.53 | 0.50 |
| 2086 | 2.02 | 0.25 | 2.32 | 0.43 | 2.03 | 0.23 | 2.45 | 0.46 | 2.02 | 0.23 | 2.53 | 0.50 |
| 2092 | 2.03 | 0.25 | 2.32 | 0.43 | 2.03 | 0.23 | 2.45 | 0.46 | 2.02 | 0.23 | 2.53 | 0.50 |
| 2098 | 2.03 | 0.25 | 2.32 | 0.42 | 2.03 | 0.23 | 2.45 | 0.46 | 2.02 | 0.23 | 2.53 | 0.49 |
| 2104 | 2.03 | 0.24 | 2.32 | 0.42 | 2.03 | 0.23 | 2.45 | 0.45 | 2.02 | 0.22 | 2.53 | 0.49 |
| 2110 | 2.03 | 0.24 | 2.32 | 0.42 | 2.03 | 0.23 | 2.45 | 0.45 | 2.02 | 0.22 | 2.53 | 0.49 |
| 2116 | 2.03 | 0.24 | 2.32 | 0.42 | 2.03 | 0.23 | 2.45 | 0.45 | 2.03 | 0.22 | 2.53 | 0.48 |
| 2122 | 2.03 | 0.24 | 2.32 | 0.41 | 2.03 | 0.23 | 2.45 | 0.45 | 2.03 | 0.22 | 2.53 | 0.48 |
| 2128 | 2.03 | 0.24 | 2.32 | 0.41 | 2.03 | 0.22 | 2.45 | 0.44 | 2.03 | 0.22 | 2.53 | 0.47 |
| 2134 | 2.04 | 0.24 | 2.32 | 0.41 | 2.04 | 0.22 | 2.45 | 0.44 | 2.03 | 0.22 | 2.53 | 0.47 |
| 2140 | 2.04 | 0.24 | 2.32 | 0.41 | 2.04 | 0.22 | 2.45 | 0.44 | 2.03 | 0.22 | 2.53 | 0.47 |
| 2146 | 2.04 | 0.24 | 2.32 | 0.40 | 2.04 | 0.22 | 2.45 | 0.43 | 2.03 | 0.22 | 2.53 | 0.46 |
| 2152 | 2.04 | 0.24 | 2.32 | 0.40 | 2.04 | 0.22 | 2.45 | 0.43 | 2.03 | 0.22 | 2.53 | 0.46 |
| 2158 | 2.04 | 0.24 | 2.32 | 0.40 | 2.04 | 0.22 | 2.45 | 0.43 | 2.03 | 0.22 | 2.53 | 0.46 |
| 2164 | 2.04 | 0.23 | 2.32 | 0.40 | 2.04 | 0.22 | 2.44 | 0.42 | 2.03 | 0.21 | 2.53 | 0.45 |
| 2170 | 2.04 | 0.23 | 2.32 | 0.39 | 2.04 | 0.22 | 2.44 | 0.42 | 2.03 | 0.21 | 2.53 | 0.45 |
| 2176 | 2.04 | 0.23 | 2.32 | 0.39 | 2.04 | 0.22 | 2.44 | 0.42 | 2.04 | 0.21 | 2.53 | 0.45 |
| 2182 | 2.04 | 0.23 | 2.32 | 0.39 | 2.04 | 0.21 | 2.44 | 0.42 | 2.04 | 0.21 | 2.53 | 0.44 |
| 2188 | 2.05 | 0.23 | 2.32 | 0.39 | 2.04 | 0.21 | 2.44 | 0.41 | 2.04 | 0.21 | 2.53 | 0.44 |
| 2194 | 2.05 | 0.23 | 2.31 | 0.39 | 2.04 | 0.21 | 2.44 | 0.41 | 2.04 | 0.21 | 2.53 | 0.44 |
| 2200 | 2.05 | 0.23 | 2.31 | 0.38 | 2.05 | 0.21 | 2.44 | 0.41 | 2.04 | 0.21 | 2.53 | 0.43 |